\documentclass[a4paper]{llncs}
\usepackage[utf8]{inputenc}
\usepackage{lmodern}
\usepackage{relsize}
\usepackage[T1]{fontenc}
\renewcommand{\_}{\text{\textscale{0.1}{\textunderscore}}}
\usepackage{imakeidx}
\usepackage{graphicx}
\usepackage{amssymb}
\usepackage{amsfonts}
\usepackage{amsmath}
\usepackage{xspace}
\usepackage{fancyvrb}
\usepackage{pdfpages}
\usepackage{wrapfig}
\usepackage{enumitem}
\usepackage[hidelinks]{hyperref}
\usepackage{xurl}
\newif\ifPaper\Papertrue
\Paperfalse
\title{Applying Second-Order Quantifier Elimination
	    in Inspecting Gödel's Ontological Proof}
\author{Christoph Wernhard}
\institute{University of Potsdam, Germany}
\newcommand{\assign}{\mathrel{\mathop:}=}

\newcommand{\f}[1]{\mathsf{#1}}
\newcommand{\true}{\top}
\newcommand{\false}{\bot}
\newcommand{\imp}{\rightarrow}

\newcommand{\equi}{\leftrightarrow}
	    
\newcommand{\eqdef}{\; 
\raisebox{-0.1ex}[0mm]{$ \stackrel{\raisebox{-0.2ex}{\tiny 
\textnormal{def}}}{=} $}\; }

\newcommand{\pplmacro}[1]{\mathit{#1}}
\newcommand{\ppldefmacro}[1]{\mathit{#1}}

\newcommand{\pplparamplain}[1]{#1}

\newcommand{\pplparamplainsup}[2]{#1^{#2}}

\makeatletter%
\newcommand{\entrymark}[1]{}%
\newcommand\entryhead{%
\@startsection{entry}{10}{\z@}{12pt plus 2pt minus 2pt}{0pt}{}}%
\makeatother
	    
\newcommand{\pplkbBefore}
{\entryhead*{}%
\setlength{\arraycolsep}{0pt}%
\pagebreak[0]%
\begin{samepage}%
\noindent%
\rule[0.5pt]{\textwidth}{2pt}\\%
\noindent}

\newcommand{\pplkbBetween}
{\setlength{\arraycolsep}{3pt}%
\\\rule[3pt]{\textwidth}{1pt}%
\par\nopagebreak\noindent Defined as\begin{center}}

\newcommand{\pplkbAfter}{\end{center}\end{samepage}\noindent}

\newcommand{\pplkbBetweenNoDisplay}{ is defined as }
\newcommand{\pplkbAfterNoDisplay}{\end{samepage}\par\noindent}	   
	   
\newcommand{\pplkbBodyBefore}{\par\noindent where\begin{center}}
\newcommand{\pplkbBodyAfter}{\end{center}}

\newcommand{\pplIsValid}[1]{\noindent This formula is valid: $#1$\par}
\newcommand{\pplIsNotValid}[1]{\noindent This formula is not valid: $#1$\par}	    
\newcommand{\pplFailedToValidate}[1]{\noindent Failed to validate this formula: $#1$\par}
\newcommand{\pplInput}{\noindent Input: }
\newcommand{\pplResult}[1]{\noindent Result of #1: }
	   
\newcounter{def}
\newcounter{appdefa}\newcounter{appdefb}\newcounter{appdefc}
\newcounter{appdefd}\newcounter{appdefe}\newcounter{appdeff}	   
\renewcommand{\theappdefa}{A\arabic{appdefa}}
\renewcommand{\theappdefb}{B\arabic{appdefb}}
\renewcommand{\theappdefc}{C\arabic{appdefc}}
\renewcommand{\theappdefd}{D\arabic{appdefd}}
\renewcommand{\theappdefe}{E\arabic{appdefe}}
\renewcommand{\theappdeff}{F\arabic{appdeff}}	   

\makeindex
\usepackage{graphicx}
\usepackage{amssymb}

\newcommand{\defname}[1]{\emph{#1}}
\newcommand{\name}[1]{\textit{#1}}
\newcommand{\g}[1]{\mathit{#1}}
\newcommand{\quoted}[1]{\dot{#1}}

\begin{document}
%
%
  \maketitle


\newcommand\blfootnote[1]{%
  \begingroup
  \fnindent0pt%
  \renewcommand\thefootnote{}\footnote{\par\vspace{-8pt} #1}%
  \addtocounter{footnote}{-1}%
  \endgroup
}

\makeatletter%
\renewcommand\entryhead{%
\@startsection{entry}{10}{\z@}{0pt plus 2pt}{0pt}{}}%
\makeatother
  
\newcommand{\PIE}{\textit{PIE}\xspace}
\newcommand{\CMProver}{\textit{CMProver}\xspace}
\newcommand{\code}[1]{{\small \texttt{#1}}}
\newcommand{\tvar}[1]{\textit{#1}}
\newsavebox{\fmbox}
\newcommand{\origproof}[1]{\vspace{1.0pt}\hspace*{\fill}#1\hspace*{\fill}\vspace{0.8pt}}
\newcommand{\origproofshort}[1]{\vspace*{-6pt}\hspace*{\fill}#1\hspace*{\fill}\vspace{0.8pt}}
\newenvironment{outbox}
               {\begin{center}
                 \noindent\begin{lrbox}{\fmbox}\begin{minipage}{0.95\textwidth}}
               {\end{minipage}\end{lrbox}\fbox{\usebox{\fmbox}}
                 \end{center}}

\begin{abstract} In recent years, Gödel's ontological proof and variations of
  it were formalized and analyzed with automated tools in various ways. We
  supplement these analyses with a modeling in an automated environment based
  on first-order logic extended by predicate quantification. Formula macros
  are used to structure complex formulas and tasks. The analysis is presented
  as a generated type-set document where informal explanations are
  interspersed with pretty-printed formulas and outputs of reasoners for
  first-order theorem proving and second-order quantifier elimination.
  Previously unnoticed or obscured aspects and details of Gödel's proof become
  apparent. Practical application possibilities of second-order quantifier
  elimination are shown and the encountered elimination tasks may serve as
  benchmarks. \end{abstract}
%
%

\newenvironment{mycenter}[1][\topsep]
  {\setlength{\topsep}{#1}\par\kern\topsep\centering}
  {\par\kern\topsep}
\renewcommand{\pplkbBefore}
  {\entryhead*{}%
  \setlength{\arraycolsep}{0pt}%
  \pagebreak[0]%
  \begin{samepage}%
  \vspace{2pt plus 2pt}%
  \noindent}
\renewcommand{\pplkbBetween}
  {\setlength{\arraycolsep}{3pt}%
  \hspace{0.5em} is defined as\begin{mycenter}[2pt plus 2pt]}
\renewcommand{\pplkbAfter}{\end{mycenter}\end{samepage}\noindent}
\renewcommand{\pplkbBodyBefore}{\par\noindent where\vspace{-12pt}\begin{mycenter}[2pt plus 2pt]}
\renewcommand{\pplkbBodyAfter}{\end{mycenter}}

\setlength{\abovedisplayskip}{2pt plus 3pt}
\setlength{\belowdisplayskip}{2pt plus 3pt}
\setlength{\abovedisplayshortskip}{2pt plus 3pt}
\setlength{\belowdisplayshortskip}{2pt plus 3pt}

\renewcommand{\pplIsValid}[1]{\noindent \textbf{This formula is valid:} $#1$\par}
\renewcommand{\pplIsNotValid}[1]{\noindent \textbf{This formula is not valid:} $#1$\par}	    
\renewcommand{\pplFailedToValidate}[1]{\noindent \textbf{Failed to validate this formula:} $#1$\par}
\renewcommand{\pplInput}{\noindent \textbf{Input:} }
\renewcommand{\pplResult}[1]{\noindent \textbf{Result of #1:}}

\renewcommand{\theappdefa}{A-\arabic{appdefa}}
\renewcommand{\theappdefb}{B-\arabic{appdefb}}
\renewcommand{\theappdefc}{C-\arabic{appdefc}}
\renewcommand{\theappdefd}{D-\arabic{appdefd}}
\renewcommand{\theappdefe}{E-\arabic{appdefe}}
\renewcommand{\theappdeff}{F-\arabic{appdeff}}	   
  
\newcommand{\scottax}[1]{\textbf{A#1}}
\newcommand{\scottthm}[1]{\textbf{T#1}}
\newcommand{\scottcor}[1]{\textbf{#1}}
\newcommand{\scottdef}[1]{\textbf{D#1}}
\newcommand{\scottlem}[1]{\textbf{#1}}
\newcommand{\ml}[1]{\textbf{#1}}
\newcommand{\modax}[1]{\textbf{#1}}

\section{Introduction}
\label{sec-intro}

  Kurt Gödel's ontological proof is bequeathed in hand-written notes by
  himself \cite{goedel:1970} and by Dana Scott \cite{scott:1970}. Since
  transcriptions of these notes were published in 1987
  \cite{sobel:1987:goedel}, the proof was analyzed, formalized and varied in
  many different logical settings.\footnote{Recently discovered further
  sources by Gödel are presented in \cite{kanckos:lethen:2019:development}.}
  \ifPaper
    \blfootnote{Copyright \copyright\ 2021 for this paper by its author. Use
    permitted under Creative Commons License Attribution 4.0 International (CC
    BY 4.0).}
  \fi
  Books by John Howard Sobel \cite{sobel:theism} and Melvin Fitting
  \cite{fitting:god} include comprehensive discussions. Branden Fitelson, Paul
  E. Oppenheimer and Edward N. Zalta
  \cite{fitelson:zalta:2007,oppenheimer:zalta:2011} investigated various
  metaphysical arguments in an automated first-order setting based on
  \name{Prover9} \cite{prover9-mace4}. Christoph Benz\-müller and Bruno
  Woltzenlogel Paleo \cite{benzmueller:woltzenlogel:2014:goedel} initiated in
  2014 the investigation of Gödel's argument with automated systems, which led
  to a large number of follow-up studies concerning its verification with
  automated tools and the human-readable yet formal representation, e.g.,
  \cite{benzmueller:woltzenlogel:2016:success,kanckos:woltzenlogl:2017:natded,benzmueller:weber:woltzenlogel:2017:assisted,kirchner:etal:2019:metaphysics,benzmueller:2020:simplified}.
  Here we add to these lines of work an inspection of Gödel's proof in a
  logical setting that so far has not been considered for this purpose. The
  expectation is that further, previously unnoticed or obscured aspects and
  details of the proof become apparent. The used framework, \PIE (``Proving,
  Interpolating, Eliminating'') \cite{cw:2016:pie,cw:2020:pie} embeds
  automated reasoners, in particular for first-order theorem proving and
  second-order quantifier elimination, in a system for defining formula macros
  and rendering \LaTeX-formatted presentations of formula macro definitions
  and reasoner outputs. In fact, the present paper is the generated output of
  such a \PIE document.\footnote{The \PIE source of this paper is
  available at \url{http://cs.christophwernhard.com/pie}.}

  The target logic of the macros is second-order logic, or, more precisely
  classical first-order logic extended by predicate quantifiers. The macro
  layer and the formulas obtained as expansions can be strictly separated. In
  our modeling of Gödel's proof we proceed by expressing large-scale steps
  (axioms, theorems) with macros whose relationships are verified by
  invocations of embedded reasoners. In this sense our formalization of may be
  considered as \emph{semi}-automated.

  Aside of providing further material for the study of Gödel's proof, the work
  shows possibilities of applying second-order quantifier elimination in a
  practical system. It appears that the functionality of the macro mechanism
  is necessary to express nontrivial applications on the basis of first- and
  second-order logic. The elimination problems that suggested themselves in
  the course of the investigation, some of which could not be solved by the
  current version of \PIE, may be useful as benchmarks for implemented
  elimination systems.\footnote{Another recent system for second-order
  quantifier elimination on the basis of first-order logic is
  \name{DLS-Forgetter} \cite{alassaf:schmidt:forgetter:2019}, which, like the
  implementation in \PIE, is based on the \name{DLS} algorithm \cite{dls}. An
  older resolution-based system, \name{SCAN},
  \cite{scan-system-paper,scan-system} can currently be invoked via a Web
  interface.}

  The rest of the paper is structured as follows: After introducing
  preliminaries in Sect.~\ref{sec-preliminaries}, Gödel's proof in the version
  of Scott is developed in Sect.~\ref{sec-proof-main}. An approach to obtain
  the weakest sufficient precondition on the accessibility relation for
  Gödel's proof with second-order quantifier elimination is then discussed in
  Sect.~\ref{sec-special-thm-3}. Section~\ref{sec-conclusion} concludes the
  paper.
  \ifPaper
     Supplementary material is provided in the report version
     \cite{cw:2021:goedel:report} of the paper.
  \else
     Appendix~\ref{app-further-aspects} discusses further aspects of Gödel's
     proof, including modal collapse and monotheism,
     App.~\ref{app-add-special-thm3} supplements
     Sect.~\ref{sec-special-thm-3}, and App.~\ref{app-expansions} shows the
     results of macro expansion for selected tasks.

     A short version of the paper appeared as \cite{cw:2021:goedel:soqe}.
  \fi
  \section{Preliminaries}
  \label{sec-preliminaries}

  A \PIE document is a Prolog source file that contains declarative formula
  macro definitions and specifications of reasoner invocations, interspersed
  with \LaTeX\ comments in the manner of literate programming
  \cite{knuth:literate:1984}. The \PIE processor expands the formula macros,
  invokes the reasoners, and compiles a \LaTeX\ document where the formula
  macro definitions and the results of reasoner invocations are
  pretty-printed. Alternatively, the processor's functionality is accessible
  from Prolog, via the interpreter and in programs. The overall processing
  time for the present paper, including reasoner invocations and \LaTeX\
  processing to produce a PDF, is about 2.5~seconds.
  
  Formula macros without parameters can play the role of formula names.
  Expressions with macros expand into formulas of first-order logic extended
  with predicate quantifiers. Hence, some means of expression that would
  naturally be used in a higher-order logic formalization of Gödel's proof are
  not available in the expansion results. Specifically, predicates in argument
  position are not permitted and there is no abstraction mechanism to
  construct predicates from formulas. However, these higher-order features are
  in Gödel's proof actually only required with respect to specific instances
  that can be expressed in first-order logic.

  As embedded reasoners we used the first-order theorem provers \name{Prover9}
  \cite{prover9-mace4} and \name{CMProver} \cite{cw:dahn:1997,cw:2016:pie},
  the first-order model generator \name{Mace4} \cite{prover9-mace4}, and an
  implementation \cite{cw:2016:pie,cw:2020:pie} of the \name{DLS} algorithm
  \cite{dls} for second-order quantifier elimination, which is based on
  Ackermann's Lemma \cite{ackermann:35}. Reasoner outputs computed during
  processing of the \PIE document are presented with the introductory phrases
  \textbf{This formula is valid:}, \textbf{This formula is not valid:} and
  \textbf{Result of elimination:}. In addition, various methods for formula
  simplification, clausification and un-Skolemization are applied in
  preprocessing, inprocessing and for output presentation.

  The generated \LaTeX\ presentation of formulas and macro definitions bears
 some footprint inherited from the Prolog syntax that is used to write
 formulas in \PIE documents. As in Prolog, predicate and constant symbols are
 written in lower case. Macro parameters and bound logical variables that are
 to be instantiated with fresh symbols at macro expansion are printed like
 Prolog variables with a capitalized initial. Where-clauses in macro
 definitions are used to display in abstracted form auxiliary Prolog code
 executed at macro expansion.

  We write formulas of modal predicate logic as formulas of classical
  first-order logic with one additional free world variable $v$ by applying
  the standard translation from \cite[Sec.~11.4]{benthem:open} (see also
  \cite[Chap.~XII]{benthem:mlcl}), which can be defined as
  {\small
  \[
  \begin{array}{r@{\hspace{0.5em}}c@{\hspace{0.5em}}l}
  \g{ST}(P(t_1,\ldots,t_n)) & \eqdef & P(v,t_1,\ldots,t_n)\\
  \g{ST}(\lnot F) & \eqdef & \lnot \g{ST}(F)\\
  \g{ST}(F \lor G) & \eqdef & \g{ST}(F) \lor \g{ST}(G)\\
  \g{ST}(\exists x F) & \eqdef & \exists x\, (\f{e}(v,x) \land \g{ST}(F))\\
  \g{ST}(\Diamond F) & \eqdef & \exists w\, (\f{r}(v,w) \land
  \bigwedge_{i \text{ s.th. } x_i \text{ free in } F} \f{e}(w,x_i) \land \g{ST}(F)\{v \mapsto w\})\\
  \end{array}
  \]}%
  An $n$-ary predicate~$P$ in the modal logic is translated into an $n+1$-ary
  predicate, where the first argument represents a world. The binary
  predicates $\f{r}$ and $\f{e}$ are used for world accessibility and
  membership in the domain of a world. The logic operators $\land, \imp,
  \equi, \forall, \Box$ can be understood as shorthands defined in terms of
  the shown operators. As target logic we neither use a two-sorted logic nor
  encode two-sortedness explicitly with relativizer predicates. However, the
  translation of modal formulas yields formulas in which all quantifications
  are relativized by~$\f{r}$ or by~$\f{e}$, which seems to subsume the effect
  of such relativizer predicates. To express that free individual symbols are
  of sort \name{world} we use the unary predicate $\f{world}$.
  Macro~\ref{def:r:world:1}, defined below, can be used as an axiom that
  relates $\f{world}$ and $\f{r}$ as far as needed for our purposes. The
  standard translation realizes with respect to the represented modal logic
  \name{varying domain semantics} (\name{actualist notion of quantification}),
  expressed with the existence predicate $\f{e}$. \name{Constant domain
  semantics} (\name{possibilist notion of quantification}) can be achieved
  with axioms that state domain increase and decrease.
  
  As technical basis for Gödel's proof we use the presentation of Scott's
  version \cite{scott:1970} in
  \cite[Fig.~1]{benzmueller:woltzenlogel:2014:goedel}, shown here as
  Fig.~\ref{fig-orig-proof}. The identifiers \scottax{1}--\scottax{5},
  \scottthm{1}--\scottthm{3}, \scottdef{1}--\scottdef{3} and \scottcor{C} of
  the involved axioms, theorems, definitions and corollary follow
  \cite{benzmueller:woltzenlogel:2014:goedel}. In addition, Lemma~\scottlem{L}
  is taken from \cite[Fig.~1]{benzmueller:weber:woltzenlogel:2017:assisted},
  where it is appears as \scottlem{L2}.

  \begin{figure}[t]
  \centering
  \setlength{\fboxsep}{8pt}
  \fbox{\parbox{0.95\textwidth}{\small
  \begin{enumerate}[topsep=0pt]
  \item[\scottax{1}] Either a property or its negation is positive, but not both
  
  \origproof{$\forall P (\f{Pos}(\lnot P) \equi \lnot \f{Pos}(P))$}
    
  \item[\scottax{2}] A property necessarily implied by a positive property is positive
  
  \origproof{$\forall P \forall Q\, ((\f{Pos}(P) \land \Box \forall x\,
  (P(x) \imp Q(x))) \imp \f{Pos}(Q))$}

  \item[\scottthm{1}] Positive properties are possibly exemplified
  
  \origproof{$\forall P\, (\f{Pos}(P) \imp \Diamond \exists x\, P(x))$}

  \item[\scottdef{1}] A \defname{God-like} being possesses all positive properties
  
  \origproof{$\f{G}(x) \equi \forall P\, (\f{Pos}(P) \imp P(x))$}

  \item[\scottax{3}] The property of being God-like is positive
  
  \origproof{$\f{Pos}(\f{G})$}

  \item[\scottcor{C}] Possibly, God exists

  \origproofshort{$\Diamond \exists x\, \f{G}(x)$}
    
  \item[\scottax{4}] Positive properties are necessarily positive
  
  \origproof{$\forall P\, (\f{Pos}(P) \imp \Box \f{Pos}(P))$}

  \item[\scottdef{2}] An \defname{essence} of an individual is a property
    possessed by it and necessarily\ implying any of its properties
  
  \origproof{$\f{Ess}(P,x) \equi P(x) \land \forall Q\, (Q(x) \imp \Box \forall y\,
  (P(y) \imp Q(y)))$}

  \item[\scottthm{2}] Being God-like is an essence of any God-like being
  
  \origproof{$\forall x\, (\f{G}(x) \imp \f{Ess}(\f{G},x))$}

  \item[\scottdef{3}] \defname{Necessary existence} of an individual is the necessary
    exemplification of all its essences
  
  \origproofshort{$\f{NE}(x) \equi \forall P (\f{Ess}(P,x) \imp \Box \exists y\, P(y))$}

  \item[\scottax{5}] Necessary existence is a positive property
  
  \origproof{$\f{Pos}(\f{NE})$}
  
  \item[\scottlem{L}] If a God-like being exists, then necessarily a God-like
  being exists
  
  \origproof{$\exists x\, \f{G}(x) \imp \Box \exists x\, \f{G}(x)$}
    
  \item[\scottthm{3}] Necessarily, God exists
  
  \origproofshort{$\Box \exists x\, \f{G}(x)$}

  \vspace{-5pt}
  \end{enumerate}
  }}
  
  \caption{Scott's version of Gödel's ontological argument \cite{scott:1970},
  adapted from
  \cite{benzmueller:woltzenlogel:2014:goedel,benzmueller:weber:woltzenlogel:2017:assisted}.}
  \label{fig-orig-proof}
  \vspace{-7pt}
  \end{figure}

  \section{Rendering Gödel's Ontological Proof}
  \label{sec-proof-main}
  
  \subsection{Positiveness -- Proving Theorem \scottthm{1}}
  \label{sec-thm-1}

  The first two axioms in Gödel's proof, \scottax{1} and \scottax{2}, are
  about \emph{positiveness} of properties. Theorem~\scottthm{1} follows from
  them. The following macros render the left-to-right direction of~\scottax{1}
  and~\scottax{2}, respectively.
%
%
\pplkbBefore
\index{ax_1_lr(V,P)@$\ppldefmacro{ax_{1}^{\rightarrow}}(\pplparamplain{V},\pplparamplain{P})$}\refstepcounter{def}\label{def:ax:1:lr:2}%
\textbf{Macro \ref{def:ax:1:lr:2}}\hspace{0.5em} $\begin{array}{l}
\ppldefmacro{ax_{1}^{\rightarrow}}(\pplparamplain{V},\pplparamplain{P})
\end{array}
$\pplkbBetween
$\begin{array}{l}
\mathsf{world}(\pplparamplain{V}) \imp  (\mathsf{pos}(\pplparamplain{V},\pplparamplainsup{N}{\prime}) \imp  \lnot  \mathsf{pos}(\pplparamplain{V},\pplparamplainsup{P}{\prime})),
\end{array}
$\pplkbAfter
\pplkbBodyBefore
$
\begin{array}{l}\pplparamplainsup{N}{\prime}\; \assign\; \quoted{\lnot  \pplparamplain{P}},\\
\pplparamplainsup{P}{\prime}\; \assign\; \quoted{\pplparamplain{P}}.

\end{array}$\pplkbBodyAfter
%
%
  \smallskip

  \noindent \name{Is positive} is represented here by the binary predicate
  $\f{pos}$, which has a world and an individual representing a predicate as
  argument. $P'$ and $N'$ are individual constants that represent a supplied
  predicate~$P$ and its complement $\lambda v x . \lnot P(v,x)$, respectively.
  The \name{where} clause specifies that at macro expansion they are replaced
  by individual constants $\quoted{P}$ and $\quoted{\lnot P}$, available for
  each predicate symbol~$P$.
  
  Throughout this analysis, we expose the current world as macro
  parameter~$V$, which facilitates identifying proofs steps where an axiom is
  not just applied with respect to the initially given current world but to
  some other reachable world.
%
%
\pplkbBefore
\index{ax_2(V,P,Q)@$\ppldefmacro{ax_{2}}(\pplparamplain{V},\pplparamplain{P},\pplparamplain{Q})$}\refstepcounter{def}\label{def:ax:2:3}%
\textbf{Macro \ref{def:ax:2:3}}\hspace{0.5em} $\begin{array}{l}
\ppldefmacro{ax_{2}}(\pplparamplain{V},\pplparamplain{P},\pplparamplain{Q})
\end{array}
$\pplkbBetween
$\begin{array}{lllll}
\mathsf{world}(\pplparamplain{V}) &&&&\; \imp \\
(\mathsf{pos}(\pplparamplain{V},\pplparamplainsup{P}{\prime}) &&\; \land \\
\hphantom{(} \forall \pplparamplain{W} \, (\mathsf{r}(\pplparamplain{V},\pplparamplain{W}) &\; \imp \\
\hphantom{(\forall \pplparamplain{W} \, (} \forall \pplparamplain{X} \, (\mathsf{e}(\pplparamplain{W},\pplparamplain{X}) \imp  (\pplparamplain{P}(\pplparamplain{W},\pplparamplain{X}) \imp  \pplparamplain{Q}(\pplparamplain{W},\pplparamplain{X})))) &&&\; \imp \\
\hphantom{(} \mathsf{pos}(\pplparamplain{V},\pplparamplainsup{Q}{\prime})),
\end{array}
$\pplkbAfter
\pplkbBodyBefore
$
\begin{array}{l}\pplparamplainsup{P}{\prime}\; \assign\; \quoted{\pplparamplain{P}},\\
\pplparamplainsup{Q}{\prime}\; \assign\; \quoted{\pplparamplain{Q}}.

\end{array}$\pplkbBodyAfter
%
%
  \smallskip
  
  \noindent As an insight-providing intermediate step for proving \scottthm{1}
  we can now derive the following lemma using just a single instance of each
  of $\begin{array}{l}
\pplmacro{ax_{1}^{\rightarrow}}
\end{array}
$ and $\begin{array}{l}
\pplmacro{ax_{2}}
\end{array}
$, where verum ($\lambda x
  . x = x$) and falsum ($\lambda x . x \neq x$) are represented as binary
  predicates $\top$ and $\bot$, whose first argument is a world.
%
%
\pplkbBefore
\index{lemma_1(V)@$\ppldefmacro{lemma_{1}}(\pplparamplain{V})$}\refstepcounter{def}\label{def:lemma:1:1}%
\textbf{Macro \ref{def:lemma:1:1}}\hspace{0.5em} $\begin{array}{l}
\ppldefmacro{lemma_{1}}(\pplparamplain{V})
\end{array}
$\pplkbBetween
$\begin{array}{l}
\mathsf{world}(\pplparamplain{V}) \imp  \lnot  \mathsf{pos}(\pplparamplain{V},\mathsf{\quoted{\bot}}).
\end{array}
$\pplkbAfter
%
%
\pplkbBefore
\index{r_world_1@$\ppldefmacro{r\_world_{1}}$}\refstepcounter{def}\label{def:r:world:1}%
\textbf{Macro \ref{def:r:world:1}}\hspace{0.5em} $\begin{array}{l}
\ppldefmacro{r\_world_{1}}
\end{array}
$\pplkbBetweenNoDisplay
$\forall \mathit{v} \forall \mathit{w} \, (\mathsf{r}(\mathit{v},\mathit{w}) \imp  \mathsf{world}(\mathit{w})).$\pplkbAfterNoDisplay
%
%
\pplkbBefore
\index{topbot_def@$\ppldefmacro{topbot\_def}$}\refstepcounter{def}\label{def:topbot:def}%
\textbf{Macro \ref{def:topbot:def}}\hspace{0.5em} $\begin{array}{l}
\ppldefmacro{topbot\_def}
\end{array}
$\pplkbBetween
$\begin{array}{ll}
\forall \mathit{v} \forall \mathit{x} \, (\mathsf{world}(\mathit{v}) \imp  (\mathsf{\top}(\mathit{v},\mathit{x}) \equi  \mathsf{e}(\mathit{v},\mathit{x}))) &\; \land \\
\forall \mathit{v} \forall \mathit{x} \, (\mathsf{world}(\mathit{v}) \imp  (\mathsf{\bot}(\mathit{v},\mathit{x}) \equi  \lnot  \mathsf{e}(\mathit{v},\mathit{x}))).
\end{array}
$\pplkbAfter
%
%
  \noindent To express the precondition for $\begin{array}{l}
\pplmacro{lemma_{1}}
\end{array}
$ we need
  some auxiliary macros concerning $\true$ and $\false$. The following
  expresses equivalence of $\top$ and $\lambda v x . \lnot \bot(v, x)$.
%
%
\pplkbBefore
\index{topbot_equiv@$\ppldefmacro{topbot\_equiv}$}\refstepcounter{def}\label{def:topbot:equiv}%
\textbf{Macro \ref{def:topbot:equiv}}\hspace{0.5em} $\begin{array}{l}
\ppldefmacro{topbot\_equiv}
\end{array}
$\pplkbBetweenNoDisplay
$\forall \mathit{v} \forall \mathit{x} \, (\mathsf{world}(\mathit{v}) \imp  (\mathsf{\top}(\mathit{v},\mathit{x}) \equi  \lnot  \mathsf{\bot}(\mathit{v},\mathit{x}))).$\pplkbAfterNoDisplay
%
%

  \smallskip

\pplIsValid{\pplmacro{topbot\_def} \imp  \pplmacro{topbot\_equiv}.}
%
%
  \smallskip
  
  \noindent The constants $\quoted{\top}$, $\quoted{\bot}$ and $\quoted{\lnot
  \top}$ designate the individuals associated with $\top$, $\bot$ and $\lambda
  v x . \lnot \top(v, x)$, respectively. The following axiom leads from the
  equivalence expressed by Macro~\ref{def:topbot:equiv} to equality of the
  associated individuals.
%
%
\pplkbBefore
\index{topbot_equiv_equal@$\ppldefmacro{topbot\_equiv\_equal}$}\refstepcounter{def}\label{def:topbot:equiv:equal}%
\textbf{Macro \ref{def:topbot:equiv:equal}}\hspace{0.5em} $\begin{array}{l}
\ppldefmacro{topbot\_equiv\_equal}
\end{array}
$\pplkbBetweenNoDisplay
$\pplmacro{topbot\_equiv} \imp  \mathsf{\quoted{\bot}}=\mathsf{\quoted{\lnot\top}}.$\pplkbAfterNoDisplay
%
%

  \smallskip
  
  \noindent Equality is understood there with respect to first-order logic,
  not qualified by a world parameter. In
  \ifPaper
     \cite{cw:2021:goedel:report}
  \else
     App.~\ref{app-lemma-1-pre-weaker}
  \fi
  alternatives are shown, where equality is replaced by a weaker
  substitutivity property. We can now give the precondition for
  $\begin{array}{l}
\pplmacro{lemma_{1}}
\end{array}
$.
%
%
\pplkbBefore
\index{pre_lemma_1(V)@$\ppldefmacro{pre\_lemma_{1}}(\pplparamplain{V})$}\refstepcounter{def}\label{def:pre:lemma:1:1}%
\textbf{Macro \ref{def:pre:lemma:1:1}}\hspace{0.5em} $\begin{array}{l}
\ppldefmacro{pre\_lemma_{1}}(\pplparamplain{V})
\end{array}
$\pplkbBetween
$\begin{array}{ll}
\pplmacro{r\_world_{1}} &\; \land \\
\pplmacro{topbot\_def} &\; \land \\
\pplmacro{topbot\_equiv\_equal} &\; \land \\
\pplmacro{ax_{1}^{\rightarrow}}(\pplparamplain{V},\mathsf{\top}) &\; \land \\
\pplmacro{ax_{2}}(\pplparamplain{V},\mathsf{\bot},\mathsf{\top}).
\end{array}
$\pplkbAfter
\pplIsValid{\pplmacro{pre\_lemma_{1}}(\mathsf{v}) \imp  \pplmacro{lemma_{1}}(\mathsf{v}).}
%
%
  \smallskip

  \noindent \scottthm{1} can be rendered by the following macro with a
  predicate parameter.
%
%
\pplkbBefore
\index{thm_1(V,P)@$\ppldefmacro{thm_{1}}(\pplparamplain{V},\pplparamplain{P})$}\refstepcounter{def}\label{def:thm:1:2}%
\textbf{Macro \ref{def:thm:1:2}}\hspace{0.5em} $\begin{array}{l}
\ppldefmacro{thm_{1}}(\pplparamplain{V},\pplparamplain{P})
\end{array}
$\pplkbBetween
$\begin{array}{lll}
\mathsf{world}(\pplparamplain{V}) &&\; \imp \\
(\mathsf{pos}(\pplparamplain{V},\pplparamplainsup{P}{\prime}) &\; \imp \\
\hphantom{(} \exists \pplparamplain{W} \, (\mathsf{r}(\pplparamplain{V},\pplparamplain{W}) \land  \exists \pplparamplain{X} \, (\mathsf{e}(\pplparamplain{W},\pplparamplain{X}) \land  \pplparamplain{P}(\pplparamplain{W},\pplparamplain{X})))),
\end{array}
$\pplkbAfter
\pplkbBodyBefore
$
\begin{array}{l}\pplparamplainsup{P}{\prime}\; \assign\; \quoted{\pplparamplain{P}}.

\end{array}$\pplkbBodyAfter
%
%
\pplkbBefore
\index{pre_thm_1(V,P)@$\ppldefmacro{pre\_thm_{1}}(\pplparamplain{V},\pplparamplain{P})$}\refstepcounter{def}\label{def:pre:thm:1:2}%
\textbf{Macro \ref{def:pre:thm:1:2}}\hspace{0.5em} $\begin{array}{l}
\ppldefmacro{pre\_thm_{1}}(\pplparamplain{V},\pplparamplain{P})
\end{array}
$\pplkbBetweenNoDisplay
$\pplmacro{lemma_{1}}(\pplparamplain{V}) \land  \pplmacro{ax_{2}}(\pplparamplain{V},\pplparamplain{P},\mathsf{\bot}).$\pplkbAfterNoDisplay
%
%
  
  \smallskip
  
\pplIsValid{\pplmacro{pre\_thm_{1}}(\mathsf{v},\mathsf{p}) \imp  \pplmacro{thm_{1}}(\mathsf{v},\mathsf{p}).}
%
%
  \smallskip

  \noindent Instances of $\begin{array}{l}
\pplmacro{thm_{1}}(\pplparamplain{V},\pplparamplain{P})
\end{array}
$ can be proven
  for arbitrary worlds~$V$ and predicates~$P$, from the respective instance of
  the precondition $\begin{array}{l}
\pplmacro{pre\_thm_{1}}(\pplparamplain{V},\pplparamplain{P})
\end{array}
$. A further instance of
  $\begin{array}{l}
\pplmacro{ax_{2}}
\end{array}
$ -- beyond that used to prove $\begin{array}{l}
\pplmacro{lemma_{1}}
\end{array}
$ --
  is required there, with respect to~$\false$ and the given predicate~$P$.

  \subsection{Possibly, God Exists -- Proving Corollary \scottcor{C}}

  Axiom~\scottax{3} and \scottthm{1} instantiated by $\name{God-like}$
  together imply corollary~\scottcor{C}. This is rendered as follows, where
  \name{God-like} is represented by $\f{g}$.
%
%
\pplkbBefore
\index{ax_3(V)@$\ppldefmacro{ax_{3}}(\pplparamplain{V})$}\refstepcounter{def}\label{def:ax:3:1}%
\textbf{Macro \ref{def:ax:3:1}}\hspace{0.5em} $\begin{array}{l}
\ppldefmacro{ax_{3}}(\pplparamplain{V})
\end{array}
$\pplkbBetween
$\begin{array}{l}
\mathsf{world}(\pplparamplain{V}) \imp  \mathsf{pos}(\pplparamplain{V},\mathsf{\quoted{g}}).
\end{array}
$\pplkbAfter
%
%
\pplkbBefore
\index{coro(V)@$\ppldefmacro{coro}(\pplparamplain{V})$}\refstepcounter{def}\label{def:coro:1}%
\textbf{Macro \ref{def:coro:1}}\hspace{0.5em} $\begin{array}{l}
\ppldefmacro{coro}(\pplparamplain{V})
\end{array}
$\pplkbBetween
$\begin{array}{l}
\mathsf{world}(\pplparamplain{V}) \imp  \exists \pplparamplain{W} \, (\mathsf{r}(\pplparamplain{V},\pplparamplain{W}) \land  \exists \pplparamplain{X} \, (\mathsf{e}(\pplparamplain{W},\pplparamplain{X}) \land  \mathsf{g}(\pplparamplain{W},\pplparamplain{X}))).
\end{array}
$\pplkbAfter
%
%
\pplkbBefore
\index{pre_coro(V)@$\ppldefmacro{pre\_coro}(\pplparamplain{V})$}\refstepcounter{def}\label{def:pre:coro:1}%
\textbf{Macro \ref{def:pre:coro:1}}\hspace{0.5em} $\begin{array}{l}
\ppldefmacro{pre\_coro}(\pplparamplain{V})
\end{array}
$\pplkbBetweenNoDisplay
$\pplmacro{thm_{1}}(\pplparamplain{V},\mathsf{g}) \land  \pplmacro{ax_{3}}(\pplparamplain{V}).$\pplkbAfterNoDisplay
%
%

  \smallskip

\pplIsValid{\pplmacro{pre\_coro}(\mathsf{v}) \imp  \pplmacro{coro}(\mathsf{v}).}
%
%
  \smallskip

  \noindent Notice that, differently from the proofs reported in
  \cite[Fig.~2]{benzmueller:woltzenlogel:2014:goedel}, \scottcor{C}, represented here
  by $\begin{array}{l}
\pplmacro{coro}
\end{array}
$, can be proven independently from the definition of
  \name{God-like}, \scottdef{1}, which is represented here by the
  Macros~$\begin{array}{l}
\pplmacro{def_{1}^{\rightarrow}}
\end{array}
$ and~$\begin{array}{l}
\pplmacro{def_{1}^{\rightarrow \lnot}}
\end{array}
$ defined below.

  \subsection{Essence -- Proving Theorem \scottthm{2}}
  \label{sec-def-g}

  With macros $\begin{array}{l}
\pplmacro{def_{1}^{\rightarrow}}
\end{array}
$ and $\begin{array}{l}
\pplmacro{def_{1}^{\rightarrow \lnot}}
\end{array}
$, defined
  now, we represent the left-to-right direction of~\scottdef{1}. Actually,
  only this direction of \scottdef{1} is required for the proving the further
  theorems.
%
%
\pplkbBefore
\index{def_1_lr(V,X,P)@$\ppldefmacro{def_{1}^{\rightarrow}}(\pplparamplain{V},\pplparamplain{X},\pplparamplain{P})$}\refstepcounter{def}\label{def:def:1:lr:3}%
\textbf{Macro \ref{def:def:1:lr:3}}\hspace{0.5em} $\begin{array}{l}
\ppldefmacro{def_{1}^{\rightarrow}}(\pplparamplain{V},\pplparamplain{X},\pplparamplain{P})
\end{array}
$\pplkbBetween
$\begin{array}{l}
\mathsf{g}(\pplparamplain{V},\pplparamplain{X}) \imp  (\mathsf{pos}(\pplparamplain{V},\pplparamplainsup{P}{\prime}) \imp  \pplparamplain{P}(\pplparamplain{V},\pplparamplain{X})),
\end{array}
$\pplkbAfter
\pplkbBodyBefore
$
\begin{array}{l}\pplparamplainsup{P}{\prime}\; \assign\; \quoted{\pplparamplain{P}}.

\end{array}$\pplkbBodyAfter
%
%
\pplkbBefore
\index{def_1_lrn(V,X,P)@$\ppldefmacro{def_{1}^{\rightarrow \lnot}}(\pplparamplain{V},\pplparamplain{X},\pplparamplain{P})$}\refstepcounter{def}\label{def:def:1:lrn:3}%
\textbf{Macro \ref{def:def:1:lrn:3}}\hspace{0.5em} $\begin{array}{l}
\ppldefmacro{def_{1}^{\rightarrow \lnot}}(\pplparamplain{V},\pplparamplain{X},\pplparamplain{P})
\end{array}
$\pplkbBetween
$\begin{array}{l}
\mathsf{g}(\pplparamplain{V},\pplparamplain{X}) \imp  (\mathsf{pos}(\pplparamplain{V},\pplparamplainsup{P}{\prime}) \imp  \lnot  \pplparamplain{P}(\pplparamplain{V},\pplparamplain{X})),
\end{array}
$\pplkbAfter
\pplkbBodyBefore
$
\begin{array}{l}\pplparamplainsup{P}{\prime}\; \assign\; \quoted{\lnot  \pplparamplain{P}}.

\end{array}$\pplkbBodyAfter
%
%

  The following macro $\begin{array}{l}
\pplmacro{val\_ess}
\end{array}
$ renders the definiens of the
  \name{essence of} relationship between a predicate and an individual
  in~\scottdef{2}. It is originally a formula with predicate quantification,
  but without application of a predicate to a predicate. The macro
  $\begin{array}{l}
\pplmacro{val\_ess}
\end{array}
$ exposes the universally quantified predicate as
  parameter~$Q$, permitting to use it instantiated with some specific
  predicate.
%
%
\pplkbBefore
\index{val_ess(V,P,X,Q)@$\ppldefmacro{val\_ess}(\pplparamplain{V},\pplparamplain{P},\pplparamplain{X},\pplparamplain{Q})$}\refstepcounter{def}\label{def:val:ess:4}%
\textbf{Macro \ref{def:val:ess:4}}\hspace{0.5em} $\begin{array}{l}
\ppldefmacro{val\_ess}(\pplparamplain{V},\pplparamplain{P},\pplparamplain{X},\pplparamplain{Q})
\end{array}
$\pplkbBetween
$\begin{array}{llll}
\pplparamplain{P}(\pplparamplain{V},\pplparamplain{X}) &&&\; \land \\
(\pplparamplain{Q}(\pplparamplain{V},\pplparamplain{X}) &&\; \imp \\
\hphantom{(} \forall \pplparamplain{W} \, (\mathsf{r}(\pplparamplain{V},\pplparamplain{W}) &\; \imp \\
\hphantom{(\forall \pplparamplain{W} \, (} \forall \pplparamplain{Y} \, (\mathsf{e}(\pplparamplain{W},\pplparamplain{Y}) \imp  (\pplparamplain{P}(\pplparamplain{W},\pplparamplain{Y}) \imp  \pplparamplain{Q}(\pplparamplain{W},\pplparamplain{Y}))))).
\end{array}
$\pplkbAfter
%
%
  \noindent The universally quantified version of $\begin{array}{l}
\pplmacro{val\_ess}
\end{array}
$ can be be
  expressed by prefixing a predicate quantifier upon~$Q$. Eliminating this
  second-order quantifier shows another view on \name{essence}.

  \smallskip

\pplInput $\forall \mathit{q} \, \pplmacro{val\_ess}(\mathsf{v},\mathsf{p},\mathsf{x},\mathit{q}).$\\
\pplResult{elimination}
\[\begin{array}{ll}
\mathsf{p}(\mathsf{v},\mathsf{x}) &\; \land \\
\forall \mathit{y} \forall \mathit{z} \, (\mathsf{e}(\mathit{y},\mathit{z}) \land  \mathsf{p}(\mathit{y},\mathit{z}) \land  \mathsf{r}(\mathsf{v},\mathit{y}) \imp  \mathit{y}=\mathsf{v}) &\; \land \\
\forall \mathit{y} \forall \mathit{z} \, (\mathsf{e}(\mathit{y},\mathit{z}) \land  \mathsf{p}(\mathit{y},\mathit{z}) \land  \mathsf{r}(\mathsf{v},\mathit{y}) \imp  \mathit{z}=\mathsf{x}).
\end{array}
\]
%
%
  \noindent
  We convert the elimination result ``manually'' to a more clear form and
  prove equivalence by referencing to the ``last result'' via a macro.
  
%
%
\pplkbBefore
\index{last_result@$\ppldefmacro{last\_result}$}\refstepcounter{def}\label{def:last:result}%
\textbf{Macro \ref{def:last:result}}\hspace{0.5em} $\begin{array}{l}
\ppldefmacro{last\_result}
\end{array}
$\pplkbBetweenNoDisplay
$\pplparamplain{F},$\pplkbAfterNoDisplay
\pplkbBodyBefore
$
\begin{array}{l}\mathrm{last\_ppl\_result(F)}.

\end{array}$\pplkbBodyAfter
\pplIsValid{\mathsf{p}(\mathsf{v},\mathsf{x}) \land  \forall \mathit{w} \, (\mathsf{r}(\mathsf{v},\mathit{w}) \imp  \forall \mathit{y} \, (\mathsf{e}(\mathit{w},\mathit{y}) \imp  (\mathsf{p}(\mathit{w},\mathit{y}) \imp  \mathit{w}=\mathsf{v} \land  \mathit{y}=\mathsf{x}))) \equi  \pplmacro{last\_result}.}
%
%
  \smallskip

  \noindent The following definition now renders~\scottdef{2} as definition of
  the predicate $\begin{array}{l}
\mathsf{ess}
\end{array}
$ in terms of $\begin{array}{l}
\pplmacro{val\_ess}
\end{array}
$.
%
%
\pplkbBefore
\index{def_2(V,P)@$\ppldefmacro{def_{2}}(\pplparamplain{V},\pplparamplain{P})$}\refstepcounter{def}\label{def:def:2:2}%
\textbf{Macro \ref{def:def:2:2}}\hspace{0.5em} $\begin{array}{l}
\ppldefmacro{def_{2}}(\pplparamplain{V},\pplparamplain{P})
\end{array}
$\pplkbBetween
$\begin{array}{ll}
\mathsf{world}(\pplparamplain{V}) &\; \imp \\
\forall \pplparamplain{X} \, (\mathsf{ess}(\pplparamplain{V},\pplparamplainsup{P}{\prime},\pplparamplain{X}) \equi  \forall \pplparamplain{Q} \, \pplmacro{val\_ess}(\pplparamplain{V},\pplparamplain{P},\pplparamplain{X},\pplparamplain{Q})),
\end{array}
$\pplkbAfter
\pplkbBodyBefore
$
\begin{array}{l}\pplparamplainsup{P}{\prime}\; \assign\; \quoted{\pplparamplain{P}}.

\end{array}$\pplkbBodyAfter
%
%
  \smallskip

  \ifPaper
     \noindent In \cite{cw:2021:goedel:report}
  \else
     \noindent In App.~\ref{app-add-essence}
  \fi
  it is shown that two observations
  about \name{essence} mentioned as \name{NOTE} in Scott's version
  \cite{scott:1970} of Gödel's proof can be derived in this modeling.
  
  The following two macros render the right-to-left direction of
  axioms~\scottax{1} and~\scottax{4}. The original axioms involve a
  universally quantified predicate that appears only in argument role. In the
  macros, it is represented by the parameter~$P$.
%
%
\pplkbBefore
\index{ax_1_rl(V,P)@$\ppldefmacro{ax_{1}^{\leftarrow}}(\pplparamplain{V},\pplparamplain{P})$}\refstepcounter{def}\label{def:ax:1:rl:2}%
\textbf{Macro \ref{def:ax:1:rl:2}}\hspace{0.5em} $\begin{array}{l}
\ppldefmacro{ax_{1}^{\leftarrow}}(\pplparamplain{V},\pplparamplain{P})
\end{array}
$\pplkbBetween
$\begin{array}{l}
\mathsf{world}(\pplparamplain{V}) \imp  (\lnot  \mathsf{pos}(\pplparamplain{V},\pplparamplainsup{P}{\prime}) \imp  \mathsf{pos}(\pplparamplain{V},\pplparamplainsup{N}{\prime})),
\end{array}
$\pplkbAfter
\pplkbBodyBefore
$
\begin{array}{l}\pplparamplainsup{N}{\prime}\; \assign\; \quoted{\lnot  \pplparamplain{P}},\\
\pplparamplainsup{P}{\prime}\; \assign\; \quoted{\pplparamplain{P}}.

\end{array}$\pplkbBodyAfter
%
%
\pplkbBefore
\index{ax_4(V,P)@$\ppldefmacro{ax_{4}}(\pplparamplain{V},\pplparamplain{P})$}\refstepcounter{def}\label{def:ax:4:2}%
\textbf{Macro \ref{def:ax:4:2}}\hspace{0.5em} $\begin{array}{l}
\ppldefmacro{ax_{4}}(\pplparamplain{V},\pplparamplain{P})
\end{array}
$\pplkbBetween
$\begin{array}{l}
\mathsf{world}(\pplparamplain{V}) \imp  (\mathsf{pos}(\pplparamplain{V},\pplparamplainsup{P}{\prime}) \imp  \forall \pplparamplain{W} \, (\mathsf{r}(\pplparamplain{V},\pplparamplain{W}) \imp  \mathsf{pos}(\pplparamplain{W},\pplparamplainsup{P}{\prime}))),
\end{array}
$\pplkbAfter
\pplkbBodyBefore
$
\begin{array}{l}\pplparamplainsup{P}{\prime}\; \assign\; \quoted{\pplparamplain{P}}.

\end{array}$\pplkbBodyAfter
%
%
  \smallskip

  \noindent Theorem~\scottthm{2} is rendered by the following macro with
  $\begin{array}{l}
\mathsf{ess}
\end{array}
$ unfolded, which permits expansion into a universal
  second-order formula without occurrence of a predicate in argument position.
%
%
\pplkbBefore
\index{proto_thm_2(V,X)@$\ppldefmacro{proto\_thm_{2}}(\pplparamplain{V},\pplparamplain{X})$}\refstepcounter{def}\label{def:proto:thm:2:2}%
\textbf{Macro \ref{def:proto:thm:2:2}}\hspace{0.5em} $\begin{array}{l}
\ppldefmacro{proto\_thm_{2}}(\pplparamplain{V},\pplparamplain{X})
\end{array}
$\pplkbBetween
$\begin{array}{l}
\mathsf{world}(\pplparamplain{V}) \imp  (\mathsf{e}(\pplparamplain{V},\pplparamplain{X}) \imp  (\mathsf{g}(\pplparamplain{V},\pplparamplain{X}) \imp  \forall \pplparamplain{Q} \, \pplmacro{val\_ess}(\pplparamplain{V},\mathsf{g},\pplparamplain{X},\pplparamplain{Q}))).
\end{array}
$\pplkbAfter
%
%
\pplkbBefore
\index{pre_proto_thm_2(V,X,Q)@$\ppldefmacro{pre\_proto\_thm_{2}}(\pplparamplain{V},\pplparamplain{X},\pplparamplain{Q})$}\refstepcounter{def}\label{def:pre:proto:thm:2:3}%
\textbf{Macro \ref{def:pre:proto:thm:2:3}}\hspace{0.5em} $\begin{array}{l}
\ppldefmacro{pre\_proto\_thm_{2}}(\pplparamplain{V},\pplparamplain{X},\pplparamplain{Q})
\end{array}
$\pplkbBetween
$\begin{array}{ll}
\pplmacro{ax_{1}^{\leftarrow}}(\pplparamplain{V},\pplparamplain{Q}) &\; \land \\
\forall \pplparamplain{W} \, (\mathsf{r}(\pplparamplain{V},\pplparamplain{W}) \imp  \forall \pplparamplain{X} \, (\mathsf{e}(\pplparamplain{W},\pplparamplain{X}) \imp  \pplmacro{def_{1}^{\rightarrow}}(\pplparamplain{W},\pplparamplain{X},\pplparamplain{Q}))) &\; \land \\
\pplmacro{def_{1}^{\rightarrow \lnot}}(\pplparamplain{V},\pplparamplain{X},\pplparamplain{Q}) &\; \land \\
\pplmacro{ax_{4}}(\pplparamplain{V},\pplparamplain{Q}).
\end{array}
$\pplkbAfter
%
%

  \smallskip

\pplIsValid{\forall \mathit{q} \, \exists \mathit{\quoted{q}} \exists \mathit{\quoted{\lnot q}} \, \pplmacro{pre\_proto\_thm_{2}}(\mathsf{v},\mathsf{x},\mathit{q}) \imp  \pplmacro{proto\_thm_{2}}(\mathsf{v},\mathsf{x}).}
%
%
  \smallskip

  \noindent In this implication on the left side the constants
  $\begin{array}{l}
\pplmacro{\quoted{q}}
\end{array}
$ and $\begin{array}{l}
\pplmacro{\quoted{\lnot q}}
\end{array}
$, which represent
  predicates~$q$ and~$\lambda v x.\lnot q(v, x)$ in argument positions, are
  existentially quantified.

  \subsection{Necessarily, God Exists -- Proving Theorem \scottthm{3}}

  The definiens of \name{necessary existence}, which is defined in
  Definition~\scottdef{3}, is rendered here by the following macro
  $\begin{array}{l}
\pplmacro{val\_ne}
\end{array}
$, expressed in terms of $\begin{array}{l}
\pplmacro{val\_ess}
\end{array}
$, the
  representation of the definiens of \name{essence}, to avoid the occurrence
  of a predicate representative in argument position.
%
%
\pplkbBefore
\index{val_ne(V,X)@$\ppldefmacro{val\_ne}(\pplparamplain{V},\pplparamplain{X})$}\refstepcounter{def}\label{def:val:ne:2}%
\textbf{Macro \ref{def:val:ne:2}}\hspace{0.5em} $\begin{array}{l}
\ppldefmacro{val\_ne}(\pplparamplain{V},\pplparamplain{X})
\end{array}
$\pplkbBetween
$\begin{array}{ll}
\forall \pplparamplain{P} \, (\forall \pplparamplain{Q} \, \pplmacro{val\_ess}(\pplparamplain{V},\pplparamplain{P},\pplparamplain{X},\pplparamplain{Q}) &\; \imp \\
\hphantom{\forall \pplparamplain{P} \, (} \forall \pplparamplain{W} \, (\mathsf{r}(\pplparamplain{V},\pplparamplain{W}) \imp  \exists \pplparamplain{Y} \, (\mathsf{e}(\pplparamplain{W},\pplparamplain{Y}) \land  \pplparamplain{P}(\pplparamplain{W},\pplparamplain{Y})))).
\end{array}
$\pplkbAfter
%
%
  \smallskip

  \noindent Eliminating the quantified predicates shows another view on
  necessary existence.

\pplInput $\pplmacro{val\_ne}(\mathsf{v},\mathsf{x}).$\\
\pplResult{elimination}
\[\begin{array}{l}
\forall \mathit{y} \, (\mathsf{r}(\mathsf{v},\mathit{y}) \imp  \mathit{y}=\mathsf{v}) \land  \forall \mathit{y} \, (\mathsf{r}(\mathsf{v},\mathit{y}) \imp  \mathsf{e}(\mathit{y},\mathsf{x})).
\end{array}
\]
%
%

  \noindent
  The elimination result can be brought into a more clear form.

\pplIsValid{\forall \mathit{w} \, (\mathsf{r}(\mathsf{v},\mathit{w}) \imp  \mathit{w}=\mathsf{v} \land  \mathsf{e}(\mathit{w},\mathsf{x})) \equi  \pplmacro{last\_result}.}
%
%
  
  \smallskip

  \noindent In analogy to the definition of the predicate $\begin{array}{l}
\mathsf{ess}
\end{array}
$
  in Macro~\ref{def:def:2:2} we define the predicate $\begin{array}{l}
\mathsf{ne}
\end{array}
$ in
  terms of $\begin{array}{l}
\pplmacro{val\_ne}
\end{array}
$.
%
%
\pplkbBefore
\index{def_3(V,X)@$\ppldefmacro{def_{3}}(\pplparamplain{V},\pplparamplain{X})$}\refstepcounter{def}\label{def:def:3:2}%
\textbf{Macro \ref{def:def:3:2}}\hspace{0.5em} $\begin{array}{l}
\ppldefmacro{def_{3}}(\pplparamplain{V},\pplparamplain{X})
\end{array}
$\pplkbBetween
$\begin{array}{l}
\mathsf{world}(\pplparamplain{V}) \imp  (\mathsf{e}(\pplparamplain{V},\pplparamplain{X}) \imp  (\mathsf{ne}(\pplparamplain{V},\pplparamplain{X}) \equi  \pplmacro{val\_ne}(\pplparamplain{V},\pplparamplain{X}))).
\end{array}
$\pplkbAfter
%
%
  \noindent The following formula renders a fragment of the definition of
  necessary existence on a ``shallow'' level, that is, in terms of just the
  predicates $\f{ess}$ and $\f{ne}$, without referring to their definientia
  $\begin{array}{l}
\pplmacro{val\_ess}
\end{array}
$ and $\begin{array}{l}
\pplmacro{val\_ne}
\end{array}
$.
%
%
\pplkbBefore
\index{def_3_lr(V,X,P)@$\ppldefmacro{def_{3}^{\rightarrow}}(\pplparamplain{V},\pplparamplain{X},\pplparamplain{P})$}\refstepcounter{def}\label{def:def:3:lr:3}%
\textbf{Macro \ref{def:def:3:lr:3}}\hspace{0.5em} $\begin{array}{l}
\ppldefmacro{def_{3}^{\rightarrow}}(\pplparamplain{V},\pplparamplain{X},\pplparamplain{P})
\end{array}
$\pplkbBetween
$\begin{array}{lllll}
\mathsf{world}(\pplparamplain{V}) &&&&\; \imp \\
(\mathsf{e}(\pplparamplain{V},\pplparamplain{X}) &&&\; \imp \\
\hphantom{(} (\mathsf{ne}(\pplparamplain{V},\pplparamplain{X}) &&\; \imp \\
\hphantom{((} (\mathsf{ess}(\pplparamplain{V},\pplparamplainsup{P}{\prime},\pplparamplain{X}) &\; \imp \\
\hphantom{(((} \forall \pplparamplain{W} \, (\mathsf{r}(\pplparamplain{V},\pplparamplain{W}) \imp  \exists \pplparamplain{Y} \, (\mathsf{e}(\pplparamplain{W},\pplparamplain{Y}) \land  \pplparamplain{P}(\pplparamplain{W},\pplparamplain{Y})))))),
\end{array}
$\pplkbAfter
\pplkbBodyBefore
$
\begin{array}{l}\pplparamplainsup{P}{\prime}\; \assign\; \quoted{\pplparamplain{P}}.

\end{array}$\pplkbBodyAfter
%
%
  \smallskip
  
  \noindent Correctness of $\begin{array}{l}
\pplmacro{def_{3}^{\rightarrow}}
\end{array}
$ can be established by
  showing that it follows from the definitions of $\f{ess}$ and $\f{ne}$.

  \smallskip
\pplIsValid{\pplmacro{def_{2}}(\mathsf{v},\mathsf{p}) \land  \pplmacro{def_{3}}(\mathsf{v},\mathsf{x}) \imp  \pplmacro{def_{3}^{\rightarrow}}(\mathsf{v},\mathsf{x},\mathsf{p}).}
%
%
  \medskip

  The following macro renders \scottthm{2}, in contrast to
  Macro~\ref{def:proto:thm:2:2} now expressed in terms of the predicate
  $\begin{array}{l}
\mathsf{ess}
\end{array}
$ instead of its definiens $\begin{array}{l}
\pplmacro{val\_ess}
\end{array}
$.
%
%
\pplkbBefore
\index{thm_2(V,X)@$\ppldefmacro{thm_{2}}(\pplparamplain{V},\pplparamplain{X})$}\refstepcounter{def}\label{def:thm:2:2}%
\textbf{Macro \ref{def:thm:2:2}}\hspace{0.5em} $\begin{array}{l}
\ppldefmacro{thm_{2}}(\pplparamplain{V},\pplparamplain{X})
\end{array}
$\pplkbBetween
$\begin{array}{l}
\mathsf{world}(\pplparamplain{V}) \imp  (\mathsf{e}(\pplparamplain{V},\pplparamplain{X}) \imp  (\mathsf{g}(\pplparamplain{V},\pplparamplain{X}) \imp  \mathsf{ess}(\pplparamplain{V},\mathsf{\quoted{g}},\pplparamplain{X}))).
\end{array}
$\pplkbAfter
%
%
  Axiom~\scottax{5} ($\f{pos}(\f{ne})$) is represented as follows.
%
%
\pplkbBefore
\index{ax_5(V)@$\ppldefmacro{ax_{5}}(\pplparamplain{V})$}\refstepcounter{def}\label{def:ax:5:1}%
\textbf{Macro \ref{def:ax:5:1}}\hspace{0.5em} $\begin{array}{l}
\ppldefmacro{ax_{5}}(\pplparamplain{V})
\end{array}
$\pplkbBetween
$\begin{array}{l}
\mathsf{world}(\pplparamplain{V}) \imp  \mathsf{pos}(\pplparamplain{V},\mathsf{\quoted{ne}}).
\end{array}
$\pplkbAfter
%
%
  Scott's version \cite{scott:1970} shows theorem~\scottthm{3} via the
  lemma~\scottlem{L}, rendered as follows.
%
%
\pplkbBefore
\index{lemma_2(V)@$\ppldefmacro{lemma_{2}}(\pplparamplain{V})$}\refstepcounter{def}\label{def:lemma:2:1}%
\textbf{Macro \ref{def:lemma:2:1}}\hspace{0.5em} $\begin{array}{l}
\ppldefmacro{lemma_{2}}(\pplparamplain{V})
\end{array}
$\pplkbBetween
$\begin{array}{lll}
\mathsf{world}(\pplparamplain{V}) &&\; \imp \\
(\exists \pplparamplain{X} \, (\mathsf{e}(\pplparamplain{V},\pplparamplain{X}) \land  \mathsf{g}(\pplparamplain{V},\pplparamplain{X})) &\; \imp \\
\hphantom{(} \forall \pplparamplain{W} \, (\mathsf{r}(\pplparamplain{V},\pplparamplain{W}) \imp  \exists \pplparamplain{Y} \, (\mathsf{e}(\pplparamplain{W},\pplparamplain{Y}) \land  \mathsf{g}(\pplparamplain{W},\pplparamplain{Y})))).
\end{array}
$\pplkbAfter
%
%
\pplkbBefore
\index{pre_lemma_2(V,X)@$\ppldefmacro{pre\_lemma_{2}}(\pplparamplain{V},\pplparamplain{X})$}\refstepcounter{def}\label{def:pre:lemma:2:2}%
\textbf{Macro \ref{def:pre:lemma:2:2}}\hspace{0.5em} $\begin{array}{l}
\ppldefmacro{pre\_lemma_{2}}(\pplparamplain{V},\pplparamplain{X})
\end{array}
$\pplkbBetween
$\begin{array}{ll}
\pplmacro{ax_{5}}(\pplparamplain{V}) &\; \land \\
\pplmacro{def_{1}^{\rightarrow}}(\pplparamplain{V},\pplparamplain{X},\mathsf{ne}) &\; \land \\
\pplmacro{def_{3}^{\rightarrow}}(\pplparamplain{V},\pplparamplain{X},\mathsf{g}) &\; \land \\
\pplmacro{thm_{2}}(\pplparamplain{V},\pplparamplain{X}).
\end{array}
$\pplkbAfter
\pplIsValid{\forall \mathit{v} \, (\forall \mathit{x} \, \pplmacro{pre\_lemma_{2}}(\mathit{v},\mathit{x}) \imp  \pplmacro{lemma_{2}}(\mathit{v})).}
%
%
  \medskip
  
  The following formula states theorem~\scottthm{3}, the overall result to show.
%
%
\pplkbBefore
\index{thm_3(V)@$\ppldefmacro{thm_{3}}(\pplparamplain{V})$}\refstepcounter{def}\label{def:thm:3:1}%
\textbf{Macro \ref{def:thm:3:1}}\hspace{0.5em} $\begin{array}{l}
\ppldefmacro{thm_{3}}(\pplparamplain{V})
\end{array}
$\pplkbBetween
$\begin{array}{l}
\mathsf{world}(\pplparamplain{V}) \imp  \forall \pplparamplain{W} \, (\mathsf{r}(\pplparamplain{V},\pplparamplain{W}) \imp  \exists \pplparamplain{Y} \, (\mathsf{e}(\pplparamplain{W},\pplparamplain{Y}) \land  \mathsf{g}(\pplparamplain{W},\pplparamplain{Y}))).
\end{array}
$\pplkbAfter
%
%
\pplkbBefore
\index{pre_thm_3(V)@$\ppldefmacro{pre\_thm_{3}}(\pplparamplain{V})$}\refstepcounter{def}\label{def:pre:thm:3:1}%
\textbf{Macro \ref{def:pre:thm:3:1}}\hspace{0.5em} $\begin{array}{l}
\ppldefmacro{pre\_thm_{3}}(\pplparamplain{V})
\end{array}
$\pplkbBetweenNoDisplay
$\pplmacro{r\_world_{1}} \land  \forall \mathit{v} \, \pplmacro{lemma_{2}}(\mathit{v}) \land  \pplmacro{coro}(\pplparamplain{V}).$\pplkbAfterNoDisplay
%
%
\pplkbBefore
\index{euclidean@$\ppldefmacro{euclidean}$}\refstepcounter{def}\label{def:euclidean}%
\textbf{Macro \ref{def:euclidean}}\hspace{0.5em} $\begin{array}{l}
\ppldefmacro{euclidean}
\end{array}
$\pplkbBetweenNoDisplay
$\forall \mathit{x} \forall \mathit{y} \forall \mathit{z} \, (\mathsf{r}(\mathit{x},\mathit{y}) \land  \mathsf{r}(\mathit{x},\mathit{z}) \imp  \mathsf{r}(\mathit{z},\mathit{y})).$\pplkbAfterNoDisplay
%
%
\pplkbBefore
\index{symmetric@$\ppldefmacro{symmetric}$}\refstepcounter{def}\label{def:symmetric}%
\textbf{Macro \ref{def:symmetric}}\hspace{0.5em} $\begin{array}{l}
\ppldefmacro{symmetric}
\end{array}
$\pplkbBetweenNoDisplay
$\forall \mathit{x} \forall \mathit{y} \, (\mathsf{r}(\mathit{x},\mathit{y}) \imp  \mathsf{r}(\mathit{y},\mathit{x})).$\pplkbAfterNoDisplay
%
%

  \smallskip

\pplIsValid{\pplmacro{symmetric} \lor  \pplmacro{euclidean} \imp  (\pplmacro{pre\_thm_{3}}(\mathsf{v}) \imp  \pplmacro{thm_{3}}(\mathsf{v})).}
%
%
  \smallskip
  
  \noindent As observed in \cite{benzmueller:woltzenlogel:2014:goedel},
  \scottthm{3} can not be just proven in the modal logic~\ml{S5}, but also
  in~\ml{KB}, whose accessibility relation is just constrained to be
  symmetric. We have shown here in a single statement that the proof is
  possible for a Euclidean as well as a symmetric accessibility relation by
  presupposing the disjunction of both properties.
  Precondition $\begin{array}{l}
\pplmacro{pre\_thm_{3}}
\end{array}
$ includes $\begin{array}{l}
\pplmacro{coro}
\end{array}
$
  instantiated with just the current world and $\begin{array}{l}
\pplmacro{lemma_{2}}
\end{array}
$ with a
  universal quantifier upon the world parameter. In fact, as shown now,
  using $\begin{array}{l}
\pplmacro{lemma_{2}}
\end{array}
$ there just instantiated with the current world
  would not be sufficient to derive $\begin{array}{l}
\pplmacro{thm_{3}}
\end{array}
$.

  \smallskip
\pplIsNotValid{\pplmacro{symmetric} \lor  \pplmacro{euclidean} \imp  (\pplmacro{r\_world_{1}} \land  \pplmacro{lemma_{2}}(\mathsf{v}) \land  \pplmacro{coro}(\mathsf{v}) \imp  \pplmacro{thm_{3}}(\mathsf{v})).}
%
%
  \medskip

  \ifPaper
    In \cite{cw:2021:goedel:report}
  \else
    In App.~\ref{app-further-aspects}
  \fi
  further aspects of Gödel's proof are
  modeled, in particular modal collapse and monotheism.

%
%
  \section{On Weakening the Frame Condition for Theorem~\scottthm{3}}
  \label{sec-special-thm-3}

  In the proof of $\begin{array}{l}
\pplmacro{thm_{3}}
\end{array}
$ from $\begin{array}{l}
\pplmacro{pre\_thm_{3}}
\end{array}
$ we used
  the additional frame condition $\begin{array}{l}
\pplmacro{euclidean} \lor  \pplmacro{symmetric}
\end{array}
$. The
  observation that the weaker \ml{KB} instead of \ml{S5} suffices to
  prove~\scottthm{3} was an important finding of
  \cite{benzmueller:woltzenlogel:2014:goedel}. Hence, the question arises
  whether the precondition on the accessibility relation can be weakened
  further.

  In general, the \defname{weakest sufficient condition}
  \cite{lin:snc,dls:snc,cw:2012:projcirc} of a formula~$G$ on a set~$Q$ of
  predicates within a formula~$F$ can be expressed as the second-order formula
  $\forall p_1 \ldots \forall p_n\, (F \imp G)$, where $p_1, \ldots, p_n$
  are all predicates that occur free in $F \imp G$ and are not members of $Q$.
  This formula denotes the weakest (with respect to entailment) formula~$H$ in
  which only predicates in $Q$ occur free such that $H \imp (F \imp G)$ is
  valid. Second-order quantifier elimination can be applied to this formula to
  ``compute'' a weakest sufficient condition, that is, converting it to a
  first-order formula, which, of course, is inherently not possible in all
  cases.

  For \scottthm{3}, the weakest sufficient condition on the accessibility
  relation $\f{r}$ and the domain membership relation and $\f{e}$ is the
  second-order formula \[\begin{array}{l}
\forall \mathit{g} \, \forall \mathit{v} \, (\pplmacro{pre\_thm_{3}}(\mathit{v}) \imp  \pplmacro{thm_{3}}(\mathit{v}))
\end{array}
.\] Unfortunately, elimination of the second-order quantifier
  upon~$g$ fails for this formula (at least with the current version of \PIE).
  But elimination succeeds for a simplified variant of the problem, which
  considers just propositional modal logic and combines two instances of
  Lemma~$\begin{array}{l}
\pplmacro{lemma_{2}}
\end{array}
$ with an unfolding of~\scottcor{C}. The way in
  which this simplification was obtained is outlined in
  \ifPaper
     \cite{cw:2021:goedel:report}.
  \else
    App.~\ref{app-add-special-thm3}.
  \fi
%
%
\pplkbBefore
\index{lemma_2_simp(V)@$\ppldefmacro{lemma_{2}\_simp}(\pplparamplain{V})$}\refstepcounter{def}\label{def:lemma:2:simp:1}%
\textbf{Macro \ref{def:lemma:2:simp:1}}\hspace{0.5em} $\begin{array}{l}
\ppldefmacro{lemma_{2}\_simp}(\pplparamplain{V})
\end{array}
$\pplkbBetween
$\begin{array}{l}
\mathsf{g}(\pplparamplain{V}) \imp  \forall \pplparamplain{W} \, (\mathsf{r}(\pplparamplain{V},\pplparamplain{W}) \imp  \mathsf{g}(\pplparamplain{W})).
\end{array}
$\pplkbAfter
%
%
\pplkbBefore
\index{pre_thm_3_simp_inst(V)@$\ppldefmacro{pre\_thm_{3}\_simp\_inst}(\pplparamplain{V})$}\refstepcounter{def}\label{def:pre:thm:3:simp:inst:1}%
\textbf{Macro \ref{def:pre:thm:3:simp:inst:1}}\hspace{0.5em} $\begin{array}{l}
\ppldefmacro{pre\_thm_{3}\_simp\_inst}(\pplparamplain{V})
\end{array}
$\pplkbBetween
$\begin{array}{ll}
\pplmacro{lemma_{2}\_simp}(\pplparamplain{V}) &\; \land \\
\exists \pplparamplain{W} \, (\mathsf{r}(\pplparamplain{V},\pplparamplain{W}) \land  \mathsf{g}(\pplparamplain{W}) \land  \pplmacro{lemma_{2}\_simp}(\pplparamplain{W})).
\end{array}
$\pplkbAfter
%
%
\pplkbBefore
\index{thm_3_simp(V)@$\ppldefmacro{thm_{3}\_simp}(\pplparamplain{V})$}\refstepcounter{def}\label{def:thm:3:simp:1}%
\textbf{Macro \ref{def:thm:3:simp:1}}\hspace{0.5em} $\begin{array}{l}
\ppldefmacro{thm_{3}\_simp}(\pplparamplain{V})
\end{array}
$\pplkbBetween
$\begin{array}{l}
\forall \pplparamplain{W} \, (\mathsf{r}(\pplparamplain{V},\pplparamplain{W}) \imp  \mathsf{g}(\pplparamplain{W})).
\end{array}
$\pplkbAfter
\pplIsValid{\pplmacro{euclidean} \lor  \pplmacro{symmetric} \imp  (\pplmacro{pre\_thm_{3}\_simp\_inst}(\mathsf{v}) \imp  \pplmacro{thm_{3}\_simp}(\mathsf{v})).}
%
%
  \smallskip

\pplInput $\forall \mathit{g} \, \forall \mathit{v} \, (\pplmacro{pre\_thm_{3}\_simp\_inst}(\mathit{v}) \imp  \pplmacro{thm_{3}\_simp}(\mathit{v})).$\\
\pplResult{elimination}
\[\begin{array}{l}
\forall \mathit{x} \forall \mathit{y} \forall \mathit{z} \, (\mathsf{r}(\mathit{x},\mathit{y}) \land  \mathsf{r}(\mathit{x},\mathit{z}) \imp  \mathsf{r}(\mathit{y},\mathit{x}) \lor  \mathsf{r}(\mathit{y},\mathit{z}) \lor  \mathit{x}=\mathit{y} \lor  \mathit{y}=\mathit{z}).
\end{array}
\]
%
%
  \noindent We write the resulting first-order formula in a slightly different
  form, give it a name, verify equivalence to the original form and show some
  of its properties.
%
%
\pplkbBefore
\index{frame_cond_simp@$\ppldefmacro{frame\_cond\_simp}$}\refstepcounter{def}\label{def:frame:cond:simp}%
\textbf{Macro \ref{def:frame:cond:simp}}\hspace{0.5em} $\begin{array}{l}
\ppldefmacro{frame\_cond\_simp}
\end{array}
$\pplkbBetween
$\begin{array}{l}
\forall \mathit{x} \forall \mathit{y} \forall \mathit{z} \, (\mathsf{r}(\mathit{x},\mathit{y}) \land  \mathsf{r}(\mathit{x},\mathit{z}) \land  \mathit{y}\neq \mathit{x} \land  \mathit{y}\neq \mathit{z} \imp  \mathsf{r}(\mathit{y},\mathit{x}) \lor  \mathsf{r}(\mathit{y},\mathit{z})).
\end{array}
$\pplkbAfter
\pplIsValid{\pplmacro{frame\_cond\_simp} \equi  \pplmacro{last\_result}.}
%
%
\pplkbBefore
\index{reflexive@$\ppldefmacro{reflexive}$}\refstepcounter{def}\label{def:reflexive}%
\textbf{Macro \ref{def:reflexive}}\hspace{0.5em} $\begin{array}{l}
\ppldefmacro{reflexive}
\end{array}
$\pplkbBetweenNoDisplay
$\forall \mathit{x} \, \mathsf{r}(\mathit{x},\mathit{x}).$\pplkbAfterNoDisplay
%
%

  \smallskip

\pplIsValid{\pplmacro{reflexive} \imp  (\pplmacro{symmetric} \lor  \pplmacro{euclidean} \equi  \pplmacro{frame\_cond\_simp}).}
\pplIsValid{\pplmacro{symmetric} \lor  \pplmacro{euclidean} \imp  \pplmacro{frame\_cond\_simp}.}
\pplIsNotValid{\pplmacro{frame\_cond\_simp} \imp  \pplmacro{symmetric} \lor  \pplmacro{euclidean}.}
%
%
  \smallskip

  \noindent Thus, the obtained frame condition $\begin{array}{l}
\pplmacro{frame\_cond\_simp}
\end{array}
$
  is under the assumption of reflexivity equivalent to
  $\begin{array}{l}
\pplmacro{symmetric} \lor  \pplmacro{euclidean}
\end{array}
$, and without that assumption strictly
  weaker. The following statement shows that this weaker frame condition also
  works for our original problem, proving~\scottthm{3}.

  \smallskip
\pplIsValid{\pplmacro{frame\_cond\_simp} \imp  (\pplmacro{pre\_thm_{3}}(\mathsf{v}) \imp  \pplmacro{thm_{3}}(\mathsf{v})).}
%
%
  \smallskip
  
  \noindent Hence, via the detour through elimination applied to a simplified
  problem, we have found a strictly weaker frame condition for \scottthm{3}
  than $\begin{array}{l}
\pplmacro{symmetric} \lor  \pplmacro{euclidean}
\end{array}
$, but, since elimination has just
  been performed on the second-order formula representing the simplified
  problem, we do not know whether it is \emph{the weakest} one.

  \section{Conclusion}
  \label{sec-conclusion}

  We reconstructed Gödel's ontological proof in an environment that integrates
  automated first-order theorem proving, second-order quantifier elimination,
  a formula macro mechanism and \LaTeX-based formula pretty-printing,
  supplementing a number of previous works that render Gödel's proof in other
  automated theorem proving environments. Particular observations of interest
  for the study of Gödel's proof that became apparent through our modeling
  include the following:

  \begin{enumerate} \item The presentation of the derivation of
  theorem~\scottthm{1} exhibits the few actually used instantiations of
  axioms~\scottax{1} and~\scottax{2}. The derivation is via a lemma, which
  might be useful as internal interface in the proof because it can be
  justified in alternate ways.

  \item Corollary~\scottcor{C} can be shown independently from the actual
  definition of \name{God-like} (\scottdef{1}) just on the basis of the
  assumption that \scottthm{1} applies to \name{God-like}.\footnote{This is
  also apparent in \cite[Fig.~4, line~20]{benzmueller:2020:simplified}.}

  \item In the whole proof, definition~\scottdef{1} is only used in the
  left-to-right direction.\footnote{This applies if \scottax{3} is given as in
  Scott's version, but not if it is derived from further general properties of
  \name{positive}, as in Gödel's original version and in \cite[Fig.~4,
  line~19]{benzmueller:2020:simplified}.}

  \item Second-order quantifier elimination yields first-order representations
 of\linebreak \name{essence} (definition~\scottdef{2}) and \name{necessary existence}
 (definition~\scottdef{3}).

  \item Lemma~\scottlem{L} can be derived independently from the definiens of
  \name{essence}. Here the predicate $\f{ess}$ appears in the respective
  expanded formula passed to the reasoner, but not its definiens.

  \item For the derivation of theorem~\scottthm{3} an accessibility
  relationship is sufficient that, unless reflexivity is assumed, is strictly
  weaker than the disjunction of the Euclidean property and symmetry.
  \end{enumerate}

  If non-experts in automated reasoning are addressed, the syntactical
  presentation of Gödel's argument is of particular importance
  \cite{benzmueller:woltzenlogel:2016:success}. We approached this requirement
  by means of formula macro definitions with the representation of input
  formulas by Prolog terms and \LaTeX\ pretty-printing for output formulas.

  Most automated formalizations of metaphysical arguments, e.g.,
  \cite{fitelson:zalta:2007,oppenheimer:zalta:2011,benzmueller:woltzenlogel:2014:goedel,benzmueller:woltzenlogel:2016:success,benzmueller:2020:simplified},
  seem closely tied to a particular system or combination of systems. Of
  course, processing a \PIE document similarly depends on the \PIE system with
  specific embedded reasoners. However, a system-independent view on the
  formalization is at least obtainable: The underlying target logic of the
  macro expansion is just the well-known classical first-order logic extended
  with predicate quantification. Reasoning tasks are only performed on the
  expanded formulas. The \PIE system can output these explicitly
  \ifPaper
    (see, e.g., \cite{cw:2021:goedel:report}),
  \else
    (see, e.g., App.~\ref{app-expansions}),
  \fi
  providing a low-level, but system-independent
  logical representation of the complete formalization. As a further
  beneficial aspect, such an explicit low-level formalization may prevent the
  unnoticed interaction with features of involved special logics.

  A limitation of our approach might be that there is no automated support for
  the passage from the low to the high level, i.e., folding into formula
  macros. If trust in proofs is an issue, steps in the overall workflow for
  which no proof representations are produced may be objectionable. This
  concerns macro expansion, formula normalization (see, however,
  \cite{nivelle:nf:2002}), pre- and postprocessing of formulas, and in
  particular second-order quantifier elimination, for which the creation of
  proofs seems an unexplored terrain. A practical makeshift is comparison with
  the few other elimination systems
  \cite[Sect.~4]{alassaf:schmidt:forgetter:2019}.
  
  In principle it should be possible to integrate second-order quantifier
  elimination as used here also into automated reasoning environments based on
  other paradigms, in particular the heterogeneous environments that involve
  forms of higher-order reasoning and are applied in
  \cite{benzmueller:woltzenlogel:2014:goedel,benzmueller:woltzenlogel:2016:success,benzmueller:weber:woltzenlogel:2017:assisted,benzmueller:2020:simplified}.

  Concerning second-order quantifier elimination, an issue that might be worth
  further investigation is the generalization of the method applied here
  ad-hoc to weaken the precondition on the accessibility relation: We started
  from an elimination problem that could not be solved (at least with the
  current implementation of \PIE), moved to a simpler problem and then
  verified that the solution of the simpler problem is also applicable to the
  original problem, where it does not represent the originally desired unique
  \name{weakest} sufficient condition, but nevertheless a condition that is
  weaker than the condition known before.

  \subsubsection*{Acknowledgments.}
  The author thanks Christoph Benzmüller
  and anonymous reviewers for helpful remarks. Funded by the Deutsche
  Forschungsgemeinschaft (DFG, German Research Foundation) --
  Project-ID~457292495.
  
%
%
\bibliographystyle{splncs04}
\bibliography{\jobname}

\ifPaper
  \end{document}
\fi
  
%
%
  
\clearpage  
\section*{Appendices}
\appendix

%
%
  \section{Further Aspects}
  \label{app-further-aspects}

  \subsection{Notes from Scott's Version Concerning Essence}
  \label{app-add-essence}
  
  The observations $\f{Ess}(P, x) \land \f{Ess}(Q, x) \imp \Box P=Q$ and
  $\f{Ess}(P, x) \imp \Box \forall y\, (P(y) \imp y=x)$ are stated as
  \name{NOTE} in Scott's version \cite{scott:1970} of Gödel's proof. We
  express them here with the predicate version $\begin{array}{l}
\mathsf{ess}
\end{array}
$ of
  $\name{essence}$ to facilitate their use as axioms in other statements.
%
%
\pplkbBefore
\index{note_1(V,P,Q)@$\ppldefmacro{note_{1}}(\pplparamplain{V},\pplparamplain{P},\pplparamplain{Q})$}\refstepcounter{appdefa}\label{def:note:1:3}%
\textbf{Macro \ref{def:note:1:3}}\hspace{0.5em} $\begin{array}{l}
\ppldefmacro{note_{1}}(\pplparamplain{V},\pplparamplain{P},\pplparamplain{Q})
\end{array}
$\pplkbBetween
$\begin{array}{llll}
\mathsf{world}(\pplparamplain{V}) &&&\; \imp \\
(\exists \pplparamplain{X} \, (\mathsf{ess}(\pplparamplain{V},\pplparamplainsup{P}{\prime},\pplparamplain{X}) \land  \mathsf{ess}(\pplparamplain{V},\pplparamplainsup{Q}{\prime},\pplparamplain{X})) &&\; \imp \\
\hphantom{(} \forall \pplparamplain{W} \, (\mathsf{r}(\pplparamplain{V},\pplparamplain{W}) &\; \imp \\
\hphantom{(\forall \pplparamplain{W} \, (} \forall \pplparamplain{Y} \, (\mathsf{e}(\pplparamplain{W},\pplparamplain{Y}) \imp  (\pplparamplain{P}(\pplparamplain{W},\pplparamplain{Y}) \equi  \pplparamplain{Q}(\pplparamplain{W},\pplparamplain{Y}))))),
\end{array}
$\pplkbAfter
\pplkbBodyBefore
$
\begin{array}{l}\pplparamplainsup{P}{\prime}\; \assign\; \quoted{\pplparamplain{P}},\\
\pplparamplainsup{Q}{\prime}\; \assign\; \quoted{\pplparamplain{Q}}.

\end{array}$\pplkbBodyAfter
\pplIsValid{\pplmacro{def_{2}}(\mathsf{v},\mathsf{p_{1}}) \land  \pplmacro{def_{2}}(\mathsf{v},\mathsf{p_{2}}) \imp  \pplmacro{note_{1}}(\mathsf{v},\mathsf{p_{1}},\mathsf{p_{2}}).}
%
%
\pplkbBefore
\index{note_2(V,P,X)@$\ppldefmacro{note_{2}}(\pplparamplain{V},\pplparamplain{P},\pplparamplain{X})$}\refstepcounter{appdefa}\label{def:note:2:3}%
\textbf{Macro \ref{def:note:2:3}}\hspace{0.5em} $\begin{array}{l}
\ppldefmacro{note_{2}}(\pplparamplain{V},\pplparamplain{P},\pplparamplain{X})
\end{array}
$\pplkbBetween
$\begin{array}{lll}
\mathsf{world}(\pplparamplain{V}) &&\; \imp \\
(\mathsf{ess}(\pplparamplain{V},\pplparamplainsup{P}{\prime},\pplparamplain{X}) &\; \imp \\
\hphantom{(} \forall \pplparamplain{W} \, (\mathsf{r}(\pplparamplain{V},\pplparamplain{W}) \imp  \forall \pplparamplain{Y} \, (\mathsf{e}(\pplparamplain{W},\pplparamplain{Y}) \imp  (\pplparamplain{P}(\pplparamplain{W},\pplparamplain{Y}) \imp  \pplparamplain{Y}=\pplparamplain{X})))),
\end{array}
$\pplkbAfter
\pplkbBodyBefore
$
\begin{array}{l}\pplparamplainsup{P}{\prime}\; \assign\; \quoted{\pplparamplain{P}}.

\end{array}$\pplkbBodyAfter
\pplIsValid{\pplmacro{def_{2}}(\mathsf{v},\mathsf{p}) \imp  \pplmacro{note_{2}}(\mathsf{v},\mathsf{p},\mathsf{x}).}
%
%
  \subsection{Modal Collapse}
  \label{app-collapse}
  
  A well-known objection to Gödel's theory is that it implies modal collapse
  \cite{sobel:1987:goedel}. To show this we need to relate $\f{world}$ and
  $\f{r}$ with $\begin{array}{l}
\pplmacro{r\_world}
\end{array}
$, which strengthens
  $\begin{array}{l}
\pplmacro{r\_world_{1}}
\end{array}
$ defined as Macro~\ref{def:r:world:1}.

%
%
\pplkbBefore
\index{r_world@$\ppldefmacro{r\_world}$}\refstepcounter{appdefa}\label{def:r:world}%
\textbf{Macro \ref{def:r:world}}\hspace{0.5em} $\begin{array}{l}
\ppldefmacro{r\_world}
\end{array}
$\pplkbBetween
$\begin{array}{l}
\forall \mathit{v} \forall \mathit{w} \, (\mathsf{r}(\mathit{v},\mathit{w}) \imp  \mathsf{world}(\mathit{v}) \land  \mathsf{world}(\mathit{w})).
\end{array}
$\pplkbAfter
%
%
\pplkbBefore
\index{collapse@$\ppldefmacro{collapse}$}\refstepcounter{appdefa}\label{def:collapse}%
\textbf{Macro \ref{def:collapse}}\hspace{0.5em} $\begin{array}{l}
\ppldefmacro{collapse}
\end{array}
$\pplkbBetween
$\begin{array}{l}
\forall \mathit{x} \forall \mathit{y} \, (\mathsf{r}(\mathit{x},\mathit{y}) \imp  \mathit{y}=\mathit{x}).
\end{array}
$\pplkbAfter
%
%
  \smallskip

  \noindent In our representation, modal collapse can be derived from the
  following preconditions, selected after Fitting's reconstruction
  \cite[Chapter~11, Section~8]{fitting:god} of Sobel's proof
  \cite{sobel:theism,sobel:1987:goedel}.
%
%
\pplkbBefore
\index{pre_collapse@$\ppldefmacro{pre\_collapse}$}\refstepcounter{appdefa}\label{def:pre:collapse}%
\textbf{Macro \ref{def:pre:collapse}}\hspace{0.5em} $\begin{array}{l}
\ppldefmacro{pre\_collapse}
\end{array}
$\pplkbBetween
$\begin{array}{ll}
\forall \mathit{x} \forall \mathit{v} \, \pplmacro{thm_{2}}(\mathit{v},\mathit{x}) &\; \land \\
\forall \mathit{x} \forall \mathit{v} \, \pplmacro{thm_{3}}(\mathit{v}) &\; \land \\
\forall \mathit{v} \, \pplmacro{def_{2}}(\mathit{v},\mathsf{g}) &\; \land \\
\pplmacro{r\_world} &\; \land \\
\pplmacro{reflexive}.
\end{array}
$\pplkbAfter
\pplIsValid{\pplmacro{pre\_collapse} \imp  \pplmacro{collapse}.}
%
%
  \smallskip

  \noindent In presence of \name{collapse}, the choice between frame
  conditions \name{symmetric} and \name{euclidean} (or the modal logics
  \ml{KB} and \ml{S5}) becomes immaterial, as both properties are implied by
  \name{collapse}. Also \scottax{4} is in presence of \name{collapse}
  redundant.

  \smallskip
\pplIsValid{\pplmacro{collapse} \imp  \pplmacro{symmetric} \land  \pplmacro{euclidean}.}
\pplIsValid{\pplmacro{collapse} \imp  \pplmacro{ax_{4}}(\mathsf{v},\mathsf{p}).}
%
%
  \subsection{Monotheism}
  \label{app-monotheism}
  
  In Fitting's system the proposition $\exists x \forall y\, (\f{g}(y) \equi
  y=x)$ can be derived \cite[Section~7.1]{fitting:god}. This can be proven in
  our representation from $\begin{array}{l}
\pplmacro{thm_{2}}
\end{array}
$, $\begin{array}{l}
\pplmacro{note_{2}}
\end{array}
$ and
  $\begin{array}{l}
\pplmacro{thm_{3}}
\end{array}
$ under the additional assumption of reflexivity of the
  accessibility relation. Without that assumption, it can be shown that $\Box
  \exists x \Box \forall y\, (\f{G}(y) \equi y=x)$:
%
%
\pplkbBefore
\index{pre_monotheism@$\ppldefmacro{pre\_monotheism}$}\refstepcounter{appdefa}\label{def:pre:monotheism}%
\textbf{Macro \ref{def:pre:monotheism}}\hspace{0.5em} $\begin{array}{l}
\ppldefmacro{pre\_monotheism}
\end{array}
$\pplkbBetween
$\begin{array}{ll}
\forall \mathit{x} \forall \mathit{v} \, \pplmacro{thm_{2}}(\mathit{v},\mathit{x}) &\; \land \\
\forall \mathit{x} \forall \mathit{v} \, \pplmacro{note_{2}}(\mathit{v},\mathsf{g},\mathit{x}) &\; \land \\
\forall \mathit{x} \forall \mathit{v} \, \pplmacro{thm_{3}}(\mathit{v}) &\; \land \\
\pplmacro{r\_world}.
\end{array}
$\pplkbAfter
%
%
\pplkbBefore
\index{monotheism@$\ppldefmacro{monotheism}$}\refstepcounter{appdefa}\label{def:monotheism}%
\textbf{Macro \ref{def:monotheism}}\hspace{0.5em} $\begin{array}{l}
\ppldefmacro{monotheism}
\end{array}
$\pplkbBetween
$\begin{array}{l}
\forall \mathit{v} \, \exists \mathit{x} \, (\mathsf{e}(\mathit{v},\mathit{x}) \land  \forall \mathit{y} \, (\mathsf{e}(\mathit{v},\mathit{y}) \imp  (\mathsf{g}(\mathit{v},\mathit{y}) \equi  \mathit{y}=\mathit{x}))).
\end{array}
$\pplkbAfter
\pplIsValid{\pplmacro{pre\_monotheism} \land  \pplmacro{reflexive} \imp  \pplmacro{monotheism}.}
%
%
\pplkbBefore
\index{nec_monotheism@$\ppldefmacro{nec\_monotheism}$}\refstepcounter{appdefa}\label{def:nec:monotheism}%
\textbf{Macro \ref{def:nec:monotheism}}\hspace{0.5em} $\begin{array}{l}
\ppldefmacro{nec\_monotheism}
\end{array}
$\pplkbBetween
$\begin{array}{lll}
\forall \mathit{v} \forall \mathit{w} \, (\mathsf{r}(\mathit{v},\mathit{w}) &&\; \imp \\
\hphantom{\forall \mathit{v} \forall \mathit{w} \, (} \exists \mathit{x} \, (\mathsf{e}(\mathit{w},\mathit{x}) &\; \land \\
\hphantom{\forall \mathit{v} \forall \mathit{w} \, (\exists \mathit{x} \, (} \forall \mathit{w_{1}} \, (\mathsf{r}(\mathit{w},\mathit{w_{1}}) \imp  \forall \mathit{y} \, (\mathsf{e}(\mathit{w_{1}},\mathit{y}) \imp  (\mathsf{g}(\mathit{w_{1}},\mathit{y}) \equi  \mathit{y}=\mathit{x}))))).
\end{array}
$\pplkbAfter
\pplIsValid{\pplmacro{pre\_monotheism} \imp  \pplmacro{nec\_monotheism}.}
%
%
  \section{Alternate Weaker Preconditions for $\begin{array}{l}
\pplmacro{pre\_lemma_{1}}
\end{array}
$}
  \label{app-lemma-1-pre-weaker}

  The precondition $\begin{array}{l}
\pplmacro{pre\_lemma_{1}}
\end{array}
$ used in Sect.~\ref{sec-thm-1}
  to derive $\begin{array}{l}
\pplmacro{lemma_{1}}
\end{array}
$ includes \[\begin{array}{l}
\pplmacro{topbot\_def} \land  \pplmacro{topbot\_equiv\_equal}
\end{array}
.\] The following is a weaker formula that is also
  sufficient for deriving $\begin{array}{l}
\pplmacro{lemma_{1}}
\end{array}
$:
%
%
\pplkbBefore
\index{topbot_alt_1@$\ppldefmacro{topbot\_alt_{1}}$}\refstepcounter{appdefb}\label{def:topbot:alt:1}%
\textbf{Macro \ref{def:topbot:alt:1}}\hspace{0.5em} $\begin{array}{l}
\ppldefmacro{topbot\_alt_{1}}
\end{array}
$\pplkbBetween
$\begin{array}{ll}
\forall \mathit{v} \, (\mathsf{world}(\mathit{v}) \imp  \forall \mathit{x} \, (\mathsf{e}(\mathit{v},\mathit{x}) \imp  \mathsf{\top}(\mathit{v},\mathit{x}))) &\; \land \\
\forall \mathit{v} \, (\mathsf{world}(\mathit{v}) \imp  (\mathsf{pos}(\mathit{v},\mathsf{\quoted{\bot}}) \imp  \mathsf{pos}(\mathit{v},\mathsf{\quoted{\lnot\top}}))).
\end{array}
$\pplkbAfter
\pplIsValid{\pplmacro{topbot\_def} \land  \pplmacro{topbot\_equiv\_equal} \imp  \pplmacro{topbot\_alt_{1}}.}
%
%
\pplkbBefore
\index{pre_lemma_1_drop_topbot(V)@$\ppldefmacro{pre\_lemma\_1\_drop\_topbot}(\pplparamplain{V})$}\refstepcounter{appdefb}\label{def:pre:lemma:1:drop:topbot:1}%
\textbf{Macro \ref{def:pre:lemma:1:drop:topbot:1}}\hspace{0.5em} $\begin{array}{l}
\ppldefmacro{pre\_lemma\_1\_drop\_topbot}(\pplparamplain{V})
\end{array}
$\pplkbBetween
$\begin{array}{l}
\pplparamplain{F},
\end{array}
$\pplkbAfter
\pplkbBodyBefore
$
\begin{array}{l}\pplparamplain{F} \textrm{ is like } \pplmacro{pre\_lemma_{1}}(\pplparamplain{V}) \textrm{ except }\\
\hspace*{2em} \true  \textrm{ instead of } \pplmacro{topbot\_def}\\
\hspace*{2em} \true  \textrm{ instead of } \pplmacro{topbot\_equiv\_equal}.

\end{array}$\pplkbBodyAfter
\pplIsValid{\pplmacro{topbot\_alt_{1}} \land  \pplmacro{pre\_lemma\_1\_drop\_topbot}(\mathsf{v}) \imp  \pplmacro{lemma_{1}}(\mathsf{v}).}
%
%
  \smallskip

  \noindent A third possibility to derive $\begin{array}{l}
\pplmacro{pre\_lemma_{1}}
\end{array}
$ is with
  the formula $\begin{array}{l}
\pplmacro{topbot\_alt_{2}}
\end{array}
$ defined below, which is like
  $\begin{array}{l}
\pplmacro{topbot\_alt_{1}}
\end{array}
$ except that $\top$ in the first conjunct is
  replaced by $\lnot \bot$:
%
%
\pplkbBefore
\index{topbot_alt_2@$\ppldefmacro{topbot\_alt_{2}}$}\refstepcounter{appdefb}\label{def:topbot:alt:2}%
\textbf{Macro \ref{def:topbot:alt:2}}\hspace{0.5em} $\begin{array}{l}
\ppldefmacro{topbot\_alt_{2}}
\end{array}
$\pplkbBetween
$\begin{array}{l}
\pplparamplain{F},
\end{array}
$\pplkbAfter
\pplkbBodyBefore
$
\begin{array}{l}\pplparamplain{F} \textrm{ is like } \pplmacro{topbot\_alt_{1}} \textrm{ except }\\
\hspace*{2em} \lnot  \mathsf{\bot}(\pplparamplain{V},\pplparamplain{X}) \textrm{ instead of } \mathsf{\top}(\pplparamplain{V},\pplparamplain{X}).

\end{array}$\pplkbBodyAfter
\pplIsValid{\pplmacro{topbot\_def} \land  \pplmacro{topbot\_equiv\_equal} \imp  \pplmacro{topbot\_alt_{2}}.}
\pplIsValid{\pplmacro{topbot\_alt_{2}} \land  \pplmacro{pre\_lemma\_1\_drop\_topbot}(\mathsf{v}) \imp  \pplmacro{lemma_{1}}(\mathsf{v}).}
%
%
  \section{The Way to the Simplified Elimination Task of
  Sect.~\ref{sec-special-thm-3}}
  \label{app-add-special-thm3}
  
  In Sect.~\ref{sec-special-thm-3} a simplified version of the second-order
  formula \[\begin{array}{l}
\forall \mathit{g} \, \forall \mathit{v} \, (\pplmacro{pre\_thm_{3}}(\mathit{v}) \imp  \pplmacro{thm_{3}}(\mathit{v}))
\end{array}
\] was
  used to compute a weakest sufficient condition by second-order quantifier
  elimination. The simplification was in two respects: an adaption to
  propositional modal logic and the explicit involvement of only two instances
  of $\begin{array}{l}
\pplmacro{lemma_{2}}
\end{array}
$ with an unfolding of~$\begin{array}{l}
\pplmacro{coro}
\end{array}
$. The
  adaption to propositional modal logic alone did not lead to success of
  elimination (with the current version of \PIE). The instantiation and
  unfolding was suggested by the manual inspection of a clausal tableau proof
  by \CMProver in the propositional modal logic setting.
%
%
\pplkbBefore
\index{coro_simp(V)@$\ppldefmacro{coro\_simp}(\pplparamplain{V})$}\refstepcounter{appdefc}\label{def:coro:simp:1}%
\textbf{Macro \ref{def:coro:simp:1}}\hspace{0.5em} $\begin{array}{l}
\ppldefmacro{coro\_simp}(\pplparamplain{V})
\end{array}
$\pplkbBetween
$\begin{array}{l}
\exists \pplparamplain{W} \, (\mathsf{r}(\pplparamplain{V},\pplparamplain{W}) \land  \mathsf{g}(\pplparamplain{W})).
\end{array}
$\pplkbAfter
%
%
\pplkbBefore
\index{pre_thm_3_simp(V)@$\ppldefmacro{pre\_thm_{3}\_simp}(\pplparamplain{V})$}\refstepcounter{appdefc}\label{def:pre:thm:3:simp:1}%
\textbf{Macro \ref{def:pre:thm:3:simp:1}}\hspace{0.5em} $\begin{array}{l}
\ppldefmacro{pre\_thm_{3}\_simp}(\pplparamplain{V})
\end{array}
$\pplkbBetween
$\begin{array}{l}
\forall \mathit{v} \, \pplmacro{lemma_{2}\_simp}(\mathit{v}) \land  \pplmacro{coro\_simp}(\pplparamplain{V}).
\end{array}
$\pplkbAfter
\pplIsValid{\pplmacro{symmetric} \imp  (\pplmacro{pre\_thm_{3}\_simp}(\mathsf{v}) \imp  \pplmacro{thm_{3}\_simp}(\mathsf{v})).}
%
%
  \medskip
  
  \noindent The tableau is shown in Fig.~\ref{fig:proof:simp:thm:3}.
  Actually only two instances of $\begin{array}{l}
\forall \mathit{v} \, \pplmacro{lemma_{2}\_simp}(\mathit{v})
\end{array}
$ are
  used in the proof. Formula $\begin{array}{l}
\pplmacro{pre\_thm_{3}\_simp\_inst}
\end{array}
$
  (Macro~\ref{def:pre:thm:3:simp:inst:1}) is a version of
  $\begin{array}{l}
\pplmacro{pre\_thm_{3}\_simp}
\end{array}
$ with the required two instances, the second
  one inserted into an unfolding of $\begin{array}{l}
\pplmacro{coro\_simp}
\end{array}
$.
  
  \newsavebox{\figaux}
  \savebox{\figaux}{$\begin{array}{l}
\forall \mathit{v} \, \pplmacro{lemma_{2}\_simp}(\mathit{v})
\end{array}
$}
  
  \begin{figure}[h] \centering
  \includegraphics[width=32em]{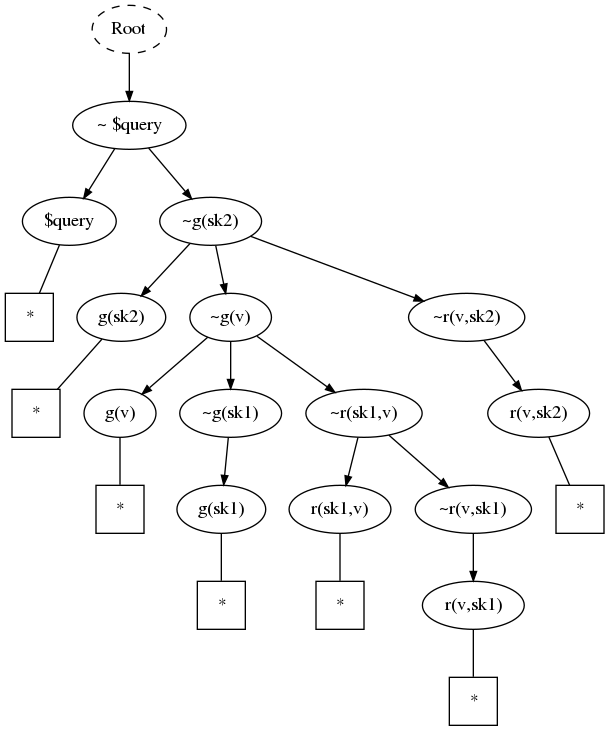} \caption{Clausal
   tableau proof of $\begin{array}{l}
\pplmacro{symmetric} \imp  (\pplmacro{pre\_thm_{3}\_simp}(\mathsf{v}) \imp  \pplmacro{thm_{3}\_simp}(\mathsf{v}))
\end{array}
$. The two instances of \usebox{\figaux} appear in the
  clausal tableau as the two ternary clauses.} \label{fig:proof:simp:thm:3}
  \end{figure}
%
%
  \clearpage
  \section{Expansions}
  \label{app-expansions}

  This appendix shows for selected proving and elimination tasks the first- or
  second-order formulas submitted after full macro expansion to reasoners.

  \smallskip
%
%
  \subsection{Formulas whose Validity was Proven by \name{Prover9} Embedded in \PIE}

  Before these expanded formulas are passed by \PIE to a first-order system,
  preprocessing is applied, which may involve the rearrangement of universal
  second-order variables into a prefix such that the validity of the resulting
  universal second-order formula equals first-order validity after
  stripping off the prefix.

  \smallskip
\noindent\textbf{Original Formula:}
\[\begin{array}{l}
\pplmacro{pre\_lemma_{1}}(\mathsf{v}) \imp  \pplmacro{lemma_{1}}(\mathsf{v}).
\end{array}
\]
\noindent\textbf{Expanded Formula:}
\[\begin{array}{lllll}
\forall \mathit{x} \forall \mathit{y} \, (\mathsf{r}(\mathit{x},\mathit{y}) \imp  \mathsf{world}(\mathit{y})) &&&\; \land \\
\forall \mathit{x} \forall \mathit{y} \, (\mathsf{world}(\mathit{x}) \imp  (\mathsf{\top}(\mathit{x},\mathit{y}) \equi  \mathsf{e}(\mathit{x},\mathit{y}))) &&&\; \land \\
\forall \mathit{x} \forall \mathit{y} \, (\mathsf{world}(\mathit{x}) \imp  (\mathsf{\bot}(\mathit{x},\mathit{y}) \equi  \lnot  \mathsf{e}(\mathit{x},\mathit{y}))) &&&\; \land \\
(\forall \mathit{x} \forall \mathit{y} \, (\mathsf{world}(\mathit{x}) \imp  (\mathsf{\top}(\mathit{x},\mathit{y}) \equi  \lnot  \mathsf{\bot}(\mathit{x},\mathit{y}))) &&\; \imp \\
\hphantom{(} \mathsf{\quoted{\bot}}=\mathsf{\quoted{\lnot\top}}) &&&\; \land \\
(\mathsf{world}(\mathsf{v}) \imp  (\mathsf{pos}(\mathsf{v},\mathsf{\quoted{\lnot\top}}) \imp  \lnot  \mathsf{pos}(\mathsf{v},\mathsf{top_{q}}))) &&&\; \land \\
(\mathsf{world}(\mathsf{v}) &&\; \imp \\
\hphantom{(} (\mathsf{pos}(\mathsf{v},\mathsf{\quoted{\bot}}) \; \land \\
\hphantom{((} \forall \mathit{x} \, (\mathsf{r}(\mathsf{v},\mathit{x}) \imp  \forall \mathit{y} \, (\mathsf{e}(\mathit{x},\mathit{y}) \imp  (\mathsf{\bot}(\mathit{x},\mathit{y}) \imp  \mathsf{\top}(\mathit{x},\mathit{y})))) &\; \imp \\
\hphantom{((} \mathsf{pos}(\mathsf{v},\mathsf{top_{q}}))) &&&&\; \imp \\
(\mathsf{world}(\mathsf{v}) \imp  \lnot  \mathsf{pos}(\mathsf{v},\mathsf{\quoted{\bot}})).
\end{array}
\]

\medskip

\noindent\textbf{Original Formula:}
\[\begin{array}{l}
\pplmacro{pre\_thm_{1}}(\mathsf{v},\mathsf{g}) \imp  \pplmacro{thm_{1}}(\mathsf{v},\mathsf{g}).
\end{array}
\]
\noindent\textbf{Expanded Formula:}
\[\begin{array}{lllll}
(\mathsf{world}(\mathsf{v}) \imp  \lnot  \mathsf{pos}(\mathsf{v},\mathsf{\quoted{\bot}})) &&&\; \land \\
(\mathsf{world}(\mathsf{v}) &&\; \imp \\
\hphantom{(} (\mathsf{pos}(\mathsf{v},\mathsf{\quoted{g}}) \; \land \\
\hphantom{((} \forall \mathit{x} \, (\mathsf{r}(\mathsf{v},\mathit{x}) \imp  \forall \mathit{y} \, (\mathsf{e}(\mathit{x},\mathit{y}) \imp  (\mathsf{g}(\mathit{x},\mathit{y}) \imp  \mathsf{\bot}(\mathit{x},\mathit{y})))) &\; \imp \\
\hphantom{((} \mathsf{pos}(\mathsf{v},\mathsf{\quoted{\bot}}))) &&&&\; \imp \\
(\mathsf{world}(\mathsf{v}) &&&\; \imp \\
\hphantom{(} (\mathsf{pos}(\mathsf{v},\mathsf{\quoted{g}}) \imp  \exists \mathit{x} \, (\mathsf{r}(\mathsf{v},\mathit{x}) \land  \exists \mathit{y} \, (\mathsf{e}(\mathit{x},\mathit{y}) \land  \mathsf{g}(\mathit{x},\mathit{y}))))).
\end{array}
\]

\medskip

\noindent\textbf{Original Formula:}
\[\begin{array}{l}
\pplmacro{pre\_coro}(\mathsf{v}) \imp  \pplmacro{coro}(\mathsf{v}).
\end{array}
\]
\noindent\textbf{Expanded Formula:}
\[\begin{array}{llll}
(\mathsf{world}(\mathsf{v}) &\; \imp \\
\hphantom{(} (\mathsf{pos}(\mathsf{v},\mathsf{\quoted{g}}) \imp  \exists \mathit{x} \, (\mathsf{r}(\mathsf{v},\mathit{x}) \land  \exists \mathit{y} \, (\mathsf{e}(\mathit{x},\mathit{y}) \land  \mathsf{g}(\mathit{x},\mathit{y}))))) &&\; \land \\
(\mathsf{world}(\mathsf{v}) \imp  \mathsf{pos}(\mathsf{v},\mathsf{\quoted{g}})) &&&\; \imp \\
(\mathsf{world}(\mathsf{v}) \imp  \exists \mathit{x} \, (\mathsf{r}(\mathsf{v},\mathit{x}) \land  \exists \mathit{y} \, (\mathsf{e}(\mathit{x},\mathit{y}) \land  \mathsf{g}(\mathit{x},\mathit{y})))).
\end{array}
\]

\medskip

\noindent\textbf{Original Formula:}
\[\begin{array}{l}
\pplmacro{pre\_proto\_thm_{2}}(\mathsf{v},\mathsf{x},\mathsf{q}) \imp  \pplmacro{proto\_thm_{2}}(\mathsf{v},\mathsf{x}).
\end{array}
\]
\noindent\textbf{Expanded Formula:}
\[\begin{array}{lllll}
\forall \mathit{p} \, \exists \mathit{y} \exists \mathit{z} \, ((\mathsf{world}(\mathsf{v}) \imp  (\lnot  \mathsf{pos}(\mathsf{v},\mathit{y}) \imp  \mathsf{pos}(\mathsf{v},\mathit{z}))) &&&\; \land \\
\hphantom{\forall \mathit{p} \, \exists \mathit{y} \exists \mathit{z} \, (} \forall \mathit{u} \, (\mathsf{r}(\mathsf{v},\mathit{u}) &&\; \imp \\
\hphantom{\forall \mathit{p} \, \exists \mathit{y} \exists \mathit{z} \, (\forall \mathit{u} \, (} \forall \mathit{w} \, (\mathsf{e}(\mathit{u},\mathit{w}) \imp  (\mathsf{g}(\mathit{u},\mathit{w}) \imp  (\mathsf{pos}(\mathit{u},\mathit{y}) \imp  \mathit{p}(\mathit{u},\mathit{w}))))) &&&\; \land \\
\hphantom{\forall \mathit{p} \, \exists \mathit{y} \exists \mathit{z} \, (} (\mathsf{g}(\mathsf{v},\mathsf{x}) \imp  (\mathsf{pos}(\mathsf{v},\mathit{z}) \imp  \lnot  \mathit{p}(\mathsf{v},\mathsf{x}))) &&&\; \land \\
\hphantom{\forall \mathit{p} \, \exists \mathit{y} \exists \mathit{z} \, (} (\mathsf{world}(\mathsf{v}) \imp  (\mathsf{pos}(\mathsf{v},\mathit{y}) \imp  \forall \mathit{u} \, (\mathsf{r}(\mathsf{v},\mathit{u}) \imp  \mathsf{pos}(\mathit{u},\mathit{y}))))) &&&&\; \imp \\
(\mathsf{world}(\mathsf{v}) &&&\; \imp \\
\hphantom{(} (\mathsf{e}(\mathsf{v},\mathsf{x}) &&\; \imp \\
\hphantom{((} (\mathsf{g}(\mathsf{v},\mathsf{x}) &\; \imp \\
\hphantom{(((} \forall \mathit{p} \, (\mathsf{g}(\mathsf{v},\mathsf{x}) \; \land \\
\hphantom{(((\forall \mathit{p} \, (} (\mathit{p}(\mathsf{v},\mathsf{x}) \; \imp \\
\hphantom{(((\forall \mathit{p} \, ((} \forall \mathit{y} \, (\mathsf{r}(\mathsf{v},\mathit{y}) \imp  \forall \mathit{z} \, (\mathsf{e}(\mathit{y},\mathit{z}) \imp  (\mathsf{g}(\mathit{y},\mathit{z}) \imp  \mathit{p}(\mathit{y},\mathit{z}))))))))).
\end{array}
\]

\medskip

\noindent\textbf{Original Formula:}
\[\begin{array}{l}
\forall \mathit{x} \, (\forall \mathit{y} \, \pplmacro{pre\_lemma_{2}}(\mathit{x},\mathit{y}) \imp  \pplmacro{lemma_{2}}(\mathit{x})).
\end{array}
\]
\noindent\textbf{Expanded Formula:}
\[\begin{array}{lllll}
\forall \mathit{x} \, (\forall \mathit{y} \, ((\mathsf{world}(\mathit{x}) \imp  \mathsf{pos}(\mathit{x},\mathsf{\quoted{ne}})) &&&\; \land \\
\hphantom{\forall \mathit{x} \, (\forall \mathit{y} \, (} (\mathsf{g}(\mathit{x},\mathit{y}) \imp  (\mathsf{pos}(\mathit{x},\mathsf{\quoted{ne}}) \imp  \mathsf{ne}(\mathit{x},\mathit{y}))) &&&\; \land \\
\hphantom{\forall \mathit{x} \, (\forall \mathit{y} \, (} (\mathsf{world}(\mathit{x}) &&\; \imp \\
\hphantom{\forall \mathit{x} \, (\forall \mathit{y} \, ((} (\mathsf{e}(\mathit{x},\mathit{y}) &\; \imp \\
\hphantom{\forall \mathit{x} \, (\forall \mathit{y} \, (((} (\mathsf{ne}(\mathit{x},\mathit{y}) \; \imp \\
\hphantom{\forall \mathit{x} \, (\forall \mathit{y} \, ((((} (\mathsf{ess}(\mathit{x},\mathsf{\quoted{g}},\mathit{y}) \; \imp \\
\hphantom{\forall \mathit{x} \, (\forall \mathit{y} \, (((((} \forall \mathit{z} \, (\mathsf{r}(\mathit{x},\mathit{z}) \imp  \exists \mathit{u} \, (\mathsf{e}(\mathit{z},\mathit{u}) \land  \mathsf{g}(\mathit{z},\mathit{u}))))))) &&&\; \land \\
\hphantom{\forall \mathit{x} \, (\forall \mathit{y} \, (} (\mathsf{world}(\mathit{x}) \imp  (\mathsf{e}(\mathit{x},\mathit{y}) \imp  (\mathsf{g}(\mathit{x},\mathit{y}) \imp  \mathsf{ess}(\mathit{x},\mathsf{\quoted{g}},\mathit{y}))))) &&&&\; \imp \\
\hphantom{\forall \mathit{x} \, (} (\mathsf{world}(\mathit{x}) &&&\; \imp \\
\hphantom{\forall \mathit{x} \, ((} (\exists \mathit{y} \, (\mathsf{e}(\mathit{x},\mathit{y}) \land  \mathsf{g}(\mathit{x},\mathit{y})) &&\; \imp \\
\hphantom{\forall \mathit{x} \, (((} \forall \mathit{y} \, (\mathsf{r}(\mathit{x},\mathit{y}) \imp  \exists \mathit{z} \, (\mathsf{e}(\mathit{y},\mathit{z}) \land  \mathsf{g}(\mathit{y},\mathit{z})))))).
\end{array}
\]

\medskip

\noindent\textbf{Original Formula:}
\[\begin{array}{l}
\pplmacro{symmetric} \lor  \pplmacro{euclidean} \imp  (\pplmacro{pre\_thm_{3}}(\mathsf{v}) \imp  \pplmacro{thm_{3}}(\mathsf{v})).
\end{array}
\]
\noindent\textbf{Expanded Formula:}
\[\begin{array}{lllll}
\forall \mathit{x} \forall \mathit{y} \, (\mathsf{r}(\mathit{x},\mathit{y}) \imp  \mathsf{r}(\mathit{y},\mathit{x})) &&&\; \lor \\
\forall \mathit{x} \forall \mathit{y} \forall \mathit{z} \, (\mathsf{r}(\mathit{x},\mathit{y}) \land  \mathsf{r}(\mathit{x},\mathit{z}) \imp  \mathsf{r}(\mathit{z},\mathit{y})) &&&&\; \imp \\
(\forall \mathit{x} \forall \mathit{y} \, (\mathsf{r}(\mathit{x},\mathit{y}) \imp  \mathsf{world}(\mathit{y})) &&\; \land \\
\hphantom{(} \forall \mathit{x} \, (\mathsf{world}(\mathit{x}) &\; \imp \\
\hphantom{(\forall \mathit{x} \, (} (\exists \mathit{y} \, (\mathsf{e}(\mathit{x},\mathit{y}) \land  \mathsf{g}(\mathit{x},\mathit{y})) \; \imp \\
\hphantom{(\forall \mathit{x} \, ((} \forall \mathit{y} \, (\mathsf{r}(\mathit{x},\mathit{y}) \imp  \exists \mathit{z} \, (\mathsf{e}(\mathit{y},\mathit{z}) \land  \mathsf{g}(\mathit{y},\mathit{z}))))) &&\; \land \\
\hphantom{(} (\mathsf{world}(\mathsf{v}) \imp  \exists \mathit{x} \, (\mathsf{r}(\mathsf{v},\mathit{x}) \land  \exists \mathit{y} \, (\mathsf{e}(\mathit{x},\mathit{y}) \land  \mathsf{g}(\mathit{x},\mathit{y})))) &&&\; \imp \\
\hphantom{(} (\mathsf{world}(\mathsf{v}) \imp  \forall \mathit{x} \, (\mathsf{r}(\mathsf{v},\mathit{x}) \imp  \exists \mathit{y} \, (\mathsf{e}(\mathit{x},\mathit{y}) \land  \mathsf{g}(\mathit{x},\mathit{y}))))).
\end{array}
\]

\medskip

\noindent\textbf{Original Formula:}
\[\begin{array}{l}
\pplmacro{frame\_cond\_simp} \imp  (\pplmacro{pre\_thm_{3}}(\mathsf{v}) \imp  \pplmacro{thm_{3}}(\mathsf{v})).
\end{array}
\]
\noindent\textbf{Expanded Formula:}
\[\begin{array}{lllll}
\forall \mathit{x} \forall \mathit{y} \forall \mathit{z} \, (\mathsf{r}(\mathit{x},\mathit{y}) \land  \mathsf{r}(\mathit{x},\mathit{z}) \land  \mathit{y}\neq \mathit{x} \land  \mathit{y}\neq \mathit{z} \imp  \mathsf{r}(\mathit{y},\mathit{x}) \lor  \mathsf{r}(\mathit{y},\mathit{z})) &&&&\; \imp \\
(\forall \mathit{x} \forall \mathit{y} \, (\mathsf{r}(\mathit{x},\mathit{y}) \imp  \mathsf{world}(\mathit{y})) &&\; \land \\
\hphantom{(} \forall \mathit{x} \, (\mathsf{world}(\mathit{x}) &\; \imp \\
\hphantom{(\forall \mathit{x} \, (} (\exists \mathit{y} \, (\mathsf{e}(\mathit{x},\mathit{y}) \land  \mathsf{g}(\mathit{x},\mathit{y})) \; \imp \\
\hphantom{(\forall \mathit{x} \, ((} \forall \mathit{y} \, (\mathsf{r}(\mathit{x},\mathit{y}) \imp  \exists \mathit{z} \, (\mathsf{e}(\mathit{y},\mathit{z}) \land  \mathsf{g}(\mathit{y},\mathit{z}))))) &&\; \land \\
\hphantom{(} (\mathsf{world}(\mathsf{v}) \imp  \exists \mathit{x} \, (\mathsf{r}(\mathsf{v},\mathit{x}) \land  \exists \mathit{y} \, (\mathsf{e}(\mathit{x},\mathit{y}) \land  \mathsf{g}(\mathit{x},\mathit{y})))) &&&\; \imp \\
\hphantom{(} (\mathsf{world}(\mathsf{v}) \imp  \forall \mathit{x} \, (\mathsf{r}(\mathsf{v},\mathit{x}) \imp  \exists \mathit{y} \, (\mathsf{e}(\mathit{x},\mathit{y}) \land  \mathsf{g}(\mathit{x},\mathit{y}))))).
\end{array}
\]

\medskip

%
%
  \subsection{Formulas whose Validity was Proven by \name{Prover9} after
  Second-Order Quantifier Elimination with \PIE}
\noindent\textbf{Original Formula:}
\[\begin{array}{l}
\pplmacro{def_{2}}(\mathsf{v},\mathsf{p}) \land  \pplmacro{def_{3}}(\mathsf{v},\mathsf{x}) \imp  \pplmacro{def_{3}^{\rightarrow}}(\mathsf{v},\mathsf{x},\mathsf{p}).
\end{array}
\]
\noindent\textbf{Expanded Formula:}
\[\begin{array}{lllll}
(\mathsf{world}(\mathsf{v}) &&\; \imp \\
\hphantom{(} \forall \mathit{y} \, (\mathsf{ess}(\mathsf{v},\mathsf{\quoted{p}},\mathit{y}) &\; \equi \\
\hphantom{(\forall \mathit{y} \, (} \forall \mathit{q} \, (\mathsf{p}(\mathsf{v},\mathit{y}) \; \land \\
\hphantom{(\forall \mathit{y} \, (\forall \mathit{q} \, (} (\mathit{q}(\mathsf{v},\mathit{y}) \; \imp \\
\hphantom{(\forall \mathit{y} \, (\forall \mathit{q} \, ((} \forall \mathit{z} \, (\mathsf{r}(\mathsf{v},\mathit{z}) \imp  \forall \mathit{u} \, (\mathsf{e}(\mathit{z},\mathit{u}) \imp  (\mathsf{p}(\mathit{z},\mathit{u}) \imp  \mathit{q}(\mathit{z},\mathit{u})))))))) &&&\; \land \\
(\mathsf{world}(\mathsf{v}) &&\; \imp \\
\hphantom{(} (\mathsf{e}(\mathsf{v},\mathsf{x}) &\; \imp \\
\hphantom{((} (\mathsf{ne}(\mathsf{v},\mathsf{x}) \; \equi \\
\hphantom{(((} \forall \mathit{q} \, (\forall \mathit{p_{1}} \, (\mathit{q}(\mathsf{v},\mathsf{x}) \; \land \\
\hphantom{(((\forall \mathit{q} \, (\forall \mathit{p_{1}} \, (} (\mathit{p_{1}}(\mathsf{v},\mathsf{x}) \; \imp \\
\hphantom{(((\forall \mathit{q} \, (\forall \mathit{p_{1}} \, ((} \forall \mathit{y} \, (\mathsf{r}(\mathsf{v},\mathit{y}) \imp  \forall \mathit{z} \, (\mathsf{e}(\mathit{y},\mathit{z}) \imp  (\mathit{q}(\mathit{y},\mathit{z}) \imp  \mathit{p_{1}}(\mathit{y},\mathit{z})))))) \; \imp \\
\hphantom{(((\forall \mathit{q} \, (} \forall \mathit{y} \, (\mathsf{r}(\mathsf{v},\mathit{y}) \imp  \exists \mathit{z} \, (\mathsf{e}(\mathit{y},\mathit{z}) \land  \mathit{q}(\mathit{y},\mathit{z}))))))) &&&&\; \imp \\
(\mathsf{world}(\mathsf{v}) &&&\; \imp \\
\hphantom{(} (\mathsf{e}(\mathsf{v},\mathsf{x}) &&\; \imp \\
\hphantom{((} (\mathsf{ne}(\mathsf{v},\mathsf{x}) &\; \imp \\
\hphantom{(((} (\mathsf{ess}(\mathsf{v},\mathsf{\quoted{p}},\mathsf{x}) \; \imp \\
\hphantom{((((} \forall \mathit{y} \, (\mathsf{r}(\mathsf{v},\mathit{y}) \imp  \exists \mathit{z} \, (\mathsf{e}(\mathit{y},\mathit{z}) \land  \mathsf{p}(\mathit{y},\mathit{z}))))))).
\end{array}
\]

\medskip

\noindent\textbf{Original Formula:}
\[\begin{array}{l}
\pplmacro{def_{2}}(\mathsf{v},\mathsf{p_{1}}) \land  \pplmacro{def_{2}}(\mathsf{v},\mathsf{p_{2}}) \imp  \pplmacro{note_{1}}(\mathsf{v},\mathsf{p_{1}},\mathsf{p_{2}}).
\end{array}
\]
\noindent\textbf{Expanded Formula:}
\[\begin{array}{lllll}
(\mathsf{world}(\mathsf{v}) &&\; \imp \\
\hphantom{(} \forall \mathit{x} \, (\mathsf{ess}(\mathsf{v},\mathsf{p1_{q}},\mathit{x}) &\; \equi \\
\hphantom{(\forall \mathit{x} \, (} \forall \mathit{p} \, (\mathsf{p_{1}}(\mathsf{v},\mathit{x}) \; \land \\
\hphantom{(\forall \mathit{x} \, (\forall \mathit{p} \, (} (\mathit{p}(\mathsf{v},\mathit{x}) \; \imp \\
\hphantom{(\forall \mathit{x} \, (\forall \mathit{p} \, ((} \forall \mathit{y} \, (\mathsf{r}(\mathsf{v},\mathit{y}) \imp  \forall \mathit{z} \, (\mathsf{e}(\mathit{y},\mathit{z}) \imp  (\mathsf{p_{1}}(\mathit{y},\mathit{z}) \imp  \mathit{p}(\mathit{y},\mathit{z})))))))) &&&\; \land \\
(\mathsf{world}(\mathsf{v}) &&\; \imp \\
\hphantom{(} \forall \mathit{x} \, (\mathsf{ess}(\mathsf{v},\mathsf{p2_{q}},\mathit{x}) &\; \equi \\
\hphantom{(\forall \mathit{x} \, (} \forall \mathit{p} \, (\mathsf{p_{2}}(\mathsf{v},\mathit{x}) \; \land \\
\hphantom{(\forall \mathit{x} \, (\forall \mathit{p} \, (} (\mathit{p}(\mathsf{v},\mathit{x}) \; \imp \\
\hphantom{(\forall \mathit{x} \, (\forall \mathit{p} \, ((} \forall \mathit{y} \, (\mathsf{r}(\mathsf{v},\mathit{y}) \imp  \forall \mathit{z} \, (\mathsf{e}(\mathit{y},\mathit{z}) \imp  (\mathsf{p_{2}}(\mathit{y},\mathit{z}) \imp  \mathit{p}(\mathit{y},\mathit{z})))))))) &&&&\; \imp \\
(\mathsf{world}(\mathsf{v}) &&&\; \imp \\
\hphantom{(} (\exists \mathit{x} \, (\mathsf{ess}(\mathsf{v},\mathsf{p1_{q}},\mathit{x}) \land  \mathsf{ess}(\mathsf{v},\mathsf{p2_{q}},\mathit{x})) &&\; \imp \\
\hphantom{((} \forall \mathit{x} \, (\mathsf{r}(\mathsf{v},\mathit{x}) \imp  \forall \mathit{y} \, (\mathsf{e}(\mathit{x},\mathit{y}) \imp  (\mathsf{p_{1}}(\mathit{x},\mathit{y}) \equi  \mathsf{p_{2}}(\mathit{x},\mathit{y})))))).
\end{array}
\]

\medskip

\noindent\textbf{Original Formula:}
\[\begin{array}{l}
\pplmacro{def_{2}}(\mathsf{v},\mathsf{p}) \imp  \pplmacro{note_{2}}(\mathsf{v},\mathsf{p},\mathsf{x}).
\end{array}
\]
\noindent\textbf{Expanded Formula:}
\[\begin{array}{lllll}
(\mathsf{world}(\mathsf{v}) &&&\; \imp \\
\hphantom{(} \forall \mathit{y} \, (\mathsf{ess}(\mathsf{v},\mathsf{\quoted{p}},\mathit{y}) &&\; \equi \\
\hphantom{(\forall \mathit{y} \, (} \forall \mathit{q} \, (\mathsf{p}(\mathsf{v},\mathit{y}) &\; \land \\
\hphantom{(\forall \mathit{y} \, (\forall \mathit{q} \, (} (\mathit{q}(\mathsf{v},\mathit{y}) \; \imp \\
\hphantom{(\forall \mathit{y} \, (\forall \mathit{q} \, ((} \forall \mathit{z} \, (\mathsf{r}(\mathsf{v},\mathit{z}) \imp  \forall \mathit{u} \, (\mathsf{e}(\mathit{z},\mathit{u}) \imp  (\mathsf{p}(\mathit{z},\mathit{u}) \imp  \mathit{q}(\mathit{z},\mathit{u})))))))) &&&&\; \imp \\
(\mathsf{world}(\mathsf{v}) &&&\; \imp \\
\hphantom{(} (\mathsf{ess}(\mathsf{v},\mathsf{\quoted{p}},\mathsf{x}) &&\; \imp \\
\hphantom{((} \forall \mathit{y} \, (\mathsf{r}(\mathsf{v},\mathit{y}) \imp  \forall \mathit{z} \, (\mathsf{e}(\mathit{y},\mathit{z}) \imp  (\mathsf{p}(\mathit{y},\mathit{z}) \imp  \mathit{z}=\mathsf{x}))))).
\end{array}
\]

\medskip

\noindent\textbf{Original Formula:}
\[\begin{array}{l}
\pplmacro{pre\_collapse} \imp  \pplmacro{collapse}.
\end{array}
\]
\noindent\textbf{Expanded Formula:}
\[\begin{array}{lllll}
\forall \mathit{x} \forall \mathit{y} \, (\mathsf{world}(\mathit{y}) \imp  (\mathsf{e}(\mathit{y},\mathit{x}) \imp  (\mathsf{g}(\mathit{y},\mathit{x}) \imp  \mathsf{ess}(\mathit{y},\mathsf{\quoted{g}},\mathit{x})))) &&&\; \land \\
\forall \mathit{x} \, (\mathsf{world}(\mathit{x}) \imp  \forall \mathit{y} \, (\mathsf{r}(\mathit{x},\mathit{y}) \imp  \exists \mathit{z} \, (\mathsf{e}(\mathit{y},\mathit{z}) \land  \mathsf{g}(\mathit{y},\mathit{z})))) &&&\; \land \\
\forall \mathit{x} \, (\mathsf{world}(\mathit{x}) &&\; \imp \\
\hphantom{\forall \mathit{x} \, (} \forall \mathit{y} \, (\mathsf{ess}(\mathit{x},\mathsf{\quoted{g}},\mathit{y}) &\; \equi \\
\hphantom{\forall \mathit{x} \, (\forall \mathit{y} \, (} \forall \mathit{p} \, (\mathsf{g}(\mathit{x},\mathit{y}) \; \land \\
\hphantom{\forall \mathit{x} \, (\forall \mathit{y} \, (\forall \mathit{p} \, (} (\mathit{p}(\mathit{x},\mathit{y}) \; \imp \\
\hphantom{\forall \mathit{x} \, (\forall \mathit{y} \, (\forall \mathit{p} \, ((} \forall \mathit{z} \, (\mathsf{r}(\mathit{x},\mathit{z}) \imp  \forall \mathit{u} \, (\mathsf{e}(\mathit{z},\mathit{u}) \imp  (\mathsf{g}(\mathit{z},\mathit{u}) \imp  \mathit{p}(\mathit{z},\mathit{u})))))))) &&&\; \land \\
\forall \mathit{x} \forall \mathit{y} \, (\mathsf{r}(\mathit{x},\mathit{y}) \imp  \mathsf{world}(\mathit{x}) \land  \mathsf{world}(\mathit{y})) &&&\; \land \\
\forall \mathit{x} \, \mathsf{r}(\mathit{x},\mathit{x}) &&&&\; \imp \\
\forall \mathit{x} \forall \mathit{y} \, (\mathsf{r}(\mathit{x},\mathit{y}) \imp  \mathit{y}=\mathit{x}).
\end{array}
\]

\medskip

%
%
  \subsection{Formulas whose Non-Validity was Proven by Mace4 Embedded in \PIE}
\noindent\textbf{Original Formula:}
\[\begin{array}{ll}
\pplmacro{symmetric} \lor  \pplmacro{euclidean} &\; \imp \\
(\pplmacro{r\_world_{1}} \land  \pplmacro{lemma_{2}}(\mathsf{v}) \land  \pplmacro{coro}(\mathsf{v}) \imp  \pplmacro{thm_{3}}(\mathsf{v})).
\end{array}
\]
\noindent\textbf{Expanded Formula:}
\[\begin{array}{lllll}
\forall \mathit{x} \forall \mathit{y} \, (\mathsf{r}(\mathit{x},\mathit{y}) \imp  \mathsf{r}(\mathit{y},\mathit{x})) &&&\; \lor \\
\forall \mathit{x} \forall \mathit{y} \forall \mathit{z} \, (\mathsf{r}(\mathit{x},\mathit{y}) \land  \mathsf{r}(\mathit{x},\mathit{z}) \imp  \mathsf{r}(\mathit{z},\mathit{y})) &&&&\; \imp \\
(\forall \mathit{x} \forall \mathit{y} \, (\mathsf{r}(\mathit{x},\mathit{y}) \imp  \mathsf{world}(\mathit{y})) &&\; \land \\
\hphantom{(} (\mathsf{world}(\mathsf{v}) &\; \imp \\
\hphantom{((} (\exists \mathit{x} \, (\mathsf{e}(\mathsf{v},\mathit{x}) \land  \mathsf{g}(\mathsf{v},\mathit{x})) \; \imp \\
\hphantom{(((} \forall \mathit{x} \, (\mathsf{r}(\mathsf{v},\mathit{x}) \imp  \exists \mathit{y} \, (\mathsf{e}(\mathit{x},\mathit{y}) \land  \mathsf{g}(\mathit{x},\mathit{y}))))) &&\; \land \\
\hphantom{(} (\mathsf{world}(\mathsf{v}) \imp  \exists \mathit{x} \, (\mathsf{r}(\mathsf{v},\mathit{x}) \land  \exists \mathit{y} \, (\mathsf{e}(\mathit{x},\mathit{y}) \land  \mathsf{g}(\mathit{x},\mathit{y})))) &&&\; \imp \\
\hphantom{(} (\mathsf{world}(\mathsf{v}) \imp  \forall \mathit{x} \, (\mathsf{r}(\mathsf{v},\mathit{x}) \imp  \exists \mathit{y} \, (\mathsf{e}(\mathit{x},\mathit{y}) \land  \mathsf{g}(\mathit{x},\mathit{y}))))).
\end{array}
\]

\medskip

%
%
  \subsection{Formulas on which Second-Order Quantifier Elimination was
  Performed with \PIE}
\noindent\textbf{Original Formula:}
\[\begin{array}{l}
\pplmacro{val\_ess}(\mathsf{v},\mathsf{p},\mathsf{x},\mathsf{q}).
\end{array}
\]
\noindent\textbf{Expanded Formula:}
\[\begin{array}{lll}
\forall \mathit{q} \, (\mathsf{p}(\mathsf{v},\mathsf{x}) &&\; \land \\
\hphantom{\forall \mathit{q} \, (} (\mathit{q}(\mathsf{v},\mathsf{x}) &\; \imp \\
\hphantom{\forall \mathit{q} \, ((} \forall \mathit{y} \, (\mathsf{r}(\mathsf{v},\mathit{y}) \imp  \forall \mathit{z} \, (\mathsf{e}(\mathit{y},\mathit{z}) \imp  (\mathsf{p}(\mathit{y},\mathit{z}) \imp  \mathit{q}(\mathit{y},\mathit{z})))))).
\end{array}
\]

\medskip

\noindent\textbf{Original Formula:}
\[\begin{array}{l}
\pplmacro{val\_ne}(\mathsf{v},\mathsf{x}).
\end{array}
\]
\noindent\textbf{Expanded Formula:}
\[\begin{array}{llll}
\forall \mathit{p} \, (\forall \mathit{q} \, (\mathit{p}(\mathsf{v},\mathsf{x}) &&\; \land \\
\hphantom{\forall \mathit{p} \, (\forall \mathit{q} \, (} (\mathit{q}(\mathsf{v},\mathsf{x}) &\; \imp \\
\hphantom{\forall \mathit{p} \, (\forall \mathit{q} \, ((} \forall \mathit{y} \, (\mathsf{r}(\mathsf{v},\mathit{y}) \imp  \forall \mathit{z} \, (\mathsf{e}(\mathit{y},\mathit{z}) \imp  (\mathit{p}(\mathit{y},\mathit{z}) \imp  \mathit{q}(\mathit{y},\mathit{z})))))) &&&\; \imp \\
\hphantom{\forall \mathit{p} \, (} \forall \mathit{y} \, (\mathsf{r}(\mathsf{v},\mathit{y}) \imp  \exists \mathit{z} \, (\mathsf{e}(\mathit{y},\mathit{z}) \land  \mathit{p}(\mathit{y},\mathit{z})))).
\end{array}
\]

\medskip

\noindent\textbf{Original Formula:}
\[\begin{array}{l}
\forall \mathit{x} \, (\pplmacro{pre\_thm_{3}\_simp\_inst}(\mathit{x}) \imp  \pplmacro{thm_{3}\_simp}(\mathit{x})).
\end{array}
\]
\noindent\textbf{Expanded Formula:}
\[\begin{array}{lll}
\forall \mathit{p} \, \forall \mathit{x} \, ((\mathit{p}(\mathit{x}) \imp  \forall \mathit{y} \, (\mathsf{r}(\mathit{x},\mathit{y}) \imp  \mathit{p}(\mathit{y}))) &\; \land \\
\hphantom{\forall \mathit{p} \, \forall \mathit{x} \, (} \exists \mathit{y} \, (\mathsf{r}(\mathit{x},\mathit{y}) \land  \mathit{p}(\mathit{y}) \land  (\mathit{p}(\mathit{y}) \imp  \forall \mathit{z} \, (\mathsf{r}(\mathit{y},\mathit{z}) \imp  \mathit{p}(\mathit{z})))) &&\; \imp \\
\hphantom{\forall \mathit{p} \, \forall \mathit{x} \, (} \forall \mathit{y} \, (\mathsf{r}(\mathit{x},\mathit{y}) \imp  \mathit{p}(\mathit{y}))).
\end{array}
\]

\medskip

%
%
  \subsection{Formulas on which Second-Order Quantifier Elimination with \PIE Failed}
\noindent\textbf{Original Formula:}
\[\begin{array}{l}
\pplmacro{pre\_thm_{3}\_simp}(\mathsf{v}) \imp  \pplmacro{thm_{3}\_simp}(\mathsf{v}).
\end{array}
\]
\noindent\textbf{Expanded Formula:}
\[\begin{array}{lll}
\forall \mathit{p} \, (\forall \mathit{x} \, (\mathit{p}(\mathit{x}) \imp  \forall \mathit{y} \, (\mathsf{r}(\mathit{x},\mathit{y}) \imp  \mathit{p}(\mathit{y}))) &\; \land \\
\hphantom{\forall \mathit{p} \, (} \exists \mathit{x} \, (\mathsf{r}(\mathsf{v},\mathit{x}) \land  \mathit{p}(\mathit{x})) &&\; \imp \\
\hphantom{\forall \mathit{p} \, (} \forall \mathit{x} \, (\mathsf{r}(\mathsf{v},\mathit{x}) \imp  \mathit{p}(\mathit{x}))).
\end{array}
\]

\medskip

\noindent\textbf{Original Formula:}
\[\begin{array}{l}
\forall \mathit{x} \, (\pplmacro{pre\_thm_{3}}(\mathit{x}) \imp  \pplmacro{thm_{3}}(\mathit{x})).
\end{array}
\]
\noindent\textbf{Expanded Formula:}
\[\begin{array}{lllll}
\forall \mathit{p} \, \forall \mathit{x} \, (\forall \mathit{y} \forall \mathit{z} \, (\mathsf{r}(\mathit{y},\mathit{z}) \imp  \mathsf{world}(\mathit{z})) &&&\; \land \\
\hphantom{\forall \mathit{p} \, \forall \mathit{x} \, (} \forall \mathit{y} \, (\mathsf{world}(\mathit{y}) &&\; \imp \\
\hphantom{\forall \mathit{p} \, \forall \mathit{x} \, (\forall \mathit{y} \, (} (\exists \mathit{z} \, (\mathsf{e}(\mathit{y},\mathit{z}) \land  \mathit{p}(\mathit{y},\mathit{z})) &\; \imp \\
\hphantom{\forall \mathit{p} \, \forall \mathit{x} \, (\forall \mathit{y} \, ((} \forall \mathit{z} \, (\mathsf{r}(\mathit{y},\mathit{z}) \imp  \exists \mathit{u} \, (\mathsf{e}(\mathit{z},\mathit{u}) \land  \mathit{p}(\mathit{z},\mathit{u}))))) &&&\; \land \\
\hphantom{\forall \mathit{p} \, \forall \mathit{x} \, (} (\mathsf{world}(\mathit{x}) \imp  \exists \mathit{y} \, (\mathsf{r}(\mathit{x},\mathit{y}) \land  \exists \mathit{z} \, (\mathsf{e}(\mathit{y},\mathit{z}) \land  \mathit{p}(\mathit{y},\mathit{z})))) &&&&\; \imp \\
\hphantom{\forall \mathit{p} \, \forall \mathit{x} \, (} (\mathsf{world}(\mathit{x}) \imp  \forall \mathit{y} \, (\mathsf{r}(\mathit{x},\mathit{y}) \imp  \exists \mathit{z} \, (\mathsf{e}(\mathit{y},\mathit{z}) \land  \mathit{p}(\mathit{y},\mathit{z}))))).
\end{array}
\]

\medskip

\end{document}